\documentclass[usenatbib,usegraphicx]{mn2e}
\usepackage{times}

\usepackage{color}

\newcommand{\dd}[1]{{\mbox{\scriptsize #1}}}
\newcommand{\ccol}[1]{\multicolumn{1}{c}{#1}}
\newcommand{\nullminus}{\makebox[0.0mm][r]{$-$}}

\title[TESELA: blank fields for astronomical observations]{TESELA: a new
Virtual Observatory tool to determine blank fields for astronomical
observations}
\author[N. Cardiel et al.]{N. Cardiel,$^{1}$\thanks{E-mail: cardiel@fis.ucm.es}
F.M.~Jim\'{e}nez-Esteban,$^{2,3,4}$ J.M.~Alacid$^{2,3}$, E.~Solano,$^{2,3}$ and
M.~Aberasturi$^{2,3}$\\
$^{1}$Departamento de Astrof\'{\i}sica y CC.\ de la Atm\'{o}sfera, Facultad de
Ciencias F\'{\i}sicas, Avenida Complutense s/n, E-28040  Madrid, Spain\\
$^{2}$Centro de Astrobiolog\'{\i}a (INTA-CSIC), Departamento de 
Astrof\'{\i}sica, PO Box 78, E-28691, Villanueva de la Ca\~nada, Madrid, Spain\\
$^{3}$Spanish Virtual Observatory, Spain\\
$^{4}$Saint Louis University, Madrid Campus, Division of Science and 
Engineering, Avenida~del~Valle 34, E-28003 Madrid, Spain}

\begin{document}

\date{Accepted \ldots Received \ldots; in original form \ldots}

\pagerange{\pageref{firstpage}--\pageref{lastpage}} \pubyear{2010}

\maketitle

\label{firstpage}

\begin{abstract}
The observation of \emph{blank fields}, regions of the sky devoid of stars down
to a given threshold magnitude, constitutes one of the typical important
calibration procedures required for the proper reduction of astronomical data
obtained in imaging mode. This work describes a method, based on the use of the
Delaunay triangulation on the surface of a sphere, that allows the easy
generation of blank fields catalogues. In addition to that, a new tool named
TESELA, accessible through the WEB, has been created to facilitate the user to
retrieve, and visualise using the VO-tool Aladin, the blank fields available
near a given position in the sky.
\end{abstract}


\section{Introduction}
\label{section:introduction}

Scientific research progress is critically based on the correct exploitation of
data obtained at the limit of the technological capabilities.  Thus prior to
data analysis and interpretation, the quality of the data treatment must be
assured in order to guarantee that the information content in that data is
preserved. In observational Astrophysics, and in particular imaging with
ground-based telescopes, data treatment is performed following data reduction
procedures. The main goal of such reduction processes is to minimize the
influence of data acquisition imperfections on the estimation of the desired
astronomical measurements (see e.g.\ \citealt{gilliland92} for a short review
on noise sources and reduction processes of CCD data). For this purpose, one
must perform appropriate manipulations of data and calibration frames,
the latter properly planned and obtained to facilitate the reduction procedure.
In this sense, image flatfielding and sky subtraction constitute two of the
most common and important reduction steps.  Their relevance should not be
underestimated, since inadequate flatfielding or sky subtraction easily leads
to the introduction of systematic uncertainties in the data. Contrary to random
errors, which can be controlled by applying normal statistical methods, the
systematic uncertainties can propagate hidden within the arithmetically
manipulated data to the final measurements, where their impact may considerably
bias their interpretation.

Concerning image flatfielding, the corresponding calibration frames needed to
proceed through this reduction step are typically divided into two categories:
images intended to provide high-frequency scale sensitivity variations (the
pixel-to-pixel response), and images whose goal is to facilitate the removal of
the two-dimensional low-frequency scale sensitivity variations of the
detectors. The former are usually
obtained from continuum lamps (either within the instrument itself or
by illuminating a flat screen or the inner dome of the telescope),
although the colour of such lamps are not expected to match the
spectral energy distributions of the science targets, and ideally dark
night sky flats, obtained by pointing the telescope to a \emph{blank
field} should be observed instead. Since this is quite expensive in
terms of observing time, this approach is very rarely used. 
The low-frequency flatfields are usually twilight exposures.
The problem here is that the time during which the sky surface brightness is
high enough to clearly overpass the signal of any star in the field of view,
but not too high to saturate the frames, are placed on two time windows every
day, one at the beginning of the evening twilight and the other at the end of
the morning twilight. The observational strategy typically consist on obtaining
a series of images, shifting the telescope position a few arcseconds between
the different exposures in order to avoid the bright stars in the field of view
to appear in the same detector pixels. Those pixels are afterwards masked when
combining the individual exposures during the reduction procedure.  In any case
very bright stars need to be avoided since the point spread function of these
objects can introduce spikes and artifacts that are not so easily removed
during the reduction of the flatfield images. For this purpose the use of
\emph{blank fields} is the best option.

Sky subtraction cannot always be performed by
measuring the night sky level in the same frame where the scientific
targets are present. This is especially true when the target
dimensions are comparable to the detector field of view. In that
circumstance separate sky frames are observed, by pointing the
telescope to a position devoid of stars as much as possible.  For the
sky subtraction procedure to be successful, it is essential that these
separate sky images are obtained under identical observing conditions,
i.e.\ very close in time and sky position,
since this is the only way to prevent varying observing conditions
that modify the sky surface brightness.

From the above discussion it is clear that the observation of sky
regions devoid of stars down to a given magnitude is a very important
aspect to be taken into account when carrying out astronomical
observations. To date, no systematic catalogue is available providing
the location in the sky of blank fields for varying limiting
magnitudes\footnote{One of the scarce resources is the web page
created by Marco Azzaro at {\tt
http://www.ing.iac.es/\~{}meteodat/blanks.htm}, where a list of 38
blank fields is available.}. With modern instrumentation intended to
be used in large imaging surveys, the problem is expected to become
more demanding. The situation is also starting to be important in
spectroscopic observations, since the advent of integral field units
with high multiplexing capabilities (and thus, increasingly larger
field of view), are expected to be one of the most common
instrumentation in ground-based observatories.

The work presented in this paper describes a method that helps to
determine the availability of blank fields in any region of the
celestial sphere, based on the use of the Delaunay triangulation. The
method has been employed to generate catalogues of blank fields to
varying limiting magnitudes. In order to facilitate the use of these
catalogues, we have created TESELA, a new tool accessible through the
WEB which provides a simple interface that allows the user to retrieve
the list of blank fields available near a given position in the sky.

Section~\ref{section:tessellating-the-sky} of this paper describes the
method and presents some statistical analysis of its application to the
\mbox{Tycho-2} \citep{Hog00a} catalogue. Section~\ref{section:TESELA}
describes the new tool TESELA. The detailed description of the
tessellation of sky subregions is presented in the Appendix.

\section{Tessellating the sky}
\label{section:tessellating-the-sky}

\subsection{The Delaunay Triangulation}
\label{subsection:delaunay-triangulation}

The Delaunay triangulation \citep{delaunay34} is a
subdivision of a geometric object (e.g.\ a surface or a volume) into a
set of simplices. A simplex, or $n$-simplex to be specific, is the
$n$-dimensional analogue of a triangle. More precisely, a simplex is
the convex hull (convex envelope) of a set of $(n+1)$ points.

In particular, for the Euclidean planar (2-dimensional) case, given a
set of points, also called nodes, the Delaunay triangulation becomes a
subdivision of the plane into triangles, whose vertices are nodes. For
each of these triangles, it is possible to determine its associated
circumcircle, the circle passing exactly through the three vertices of
the triangle, and whose center, the circumcentre, can easily be
computed as the intersection of the three perpendicular
bisectors. Interestingly, in a Delaunay triangulation all the
triangles satisfy the \emph{empty circumcircle interior property},
which states that all the circumcircles are empty, i.e., there are no
nodes inside any of the computed circumcircles.

\subsection{Applying the Delaunay triangulation to the celestial sphere}
\label{subsection:celestial-sphere}

Fortunately the Delaunay triangulation is not restricted to the
Euclidean 2-dimensional case. It can be applied to other surfaces, in
particular to the 2-dimensional surface of a 3-dimensional sphere. In
this situation, the tessellation of the spherical surface is built
with spherical triangles, i.e., triangles on the sphere whose sides
are great circles.

\begin{figure*}
\centering
\includegraphics[viewport=150 78 659 566,width=\columnwidth]{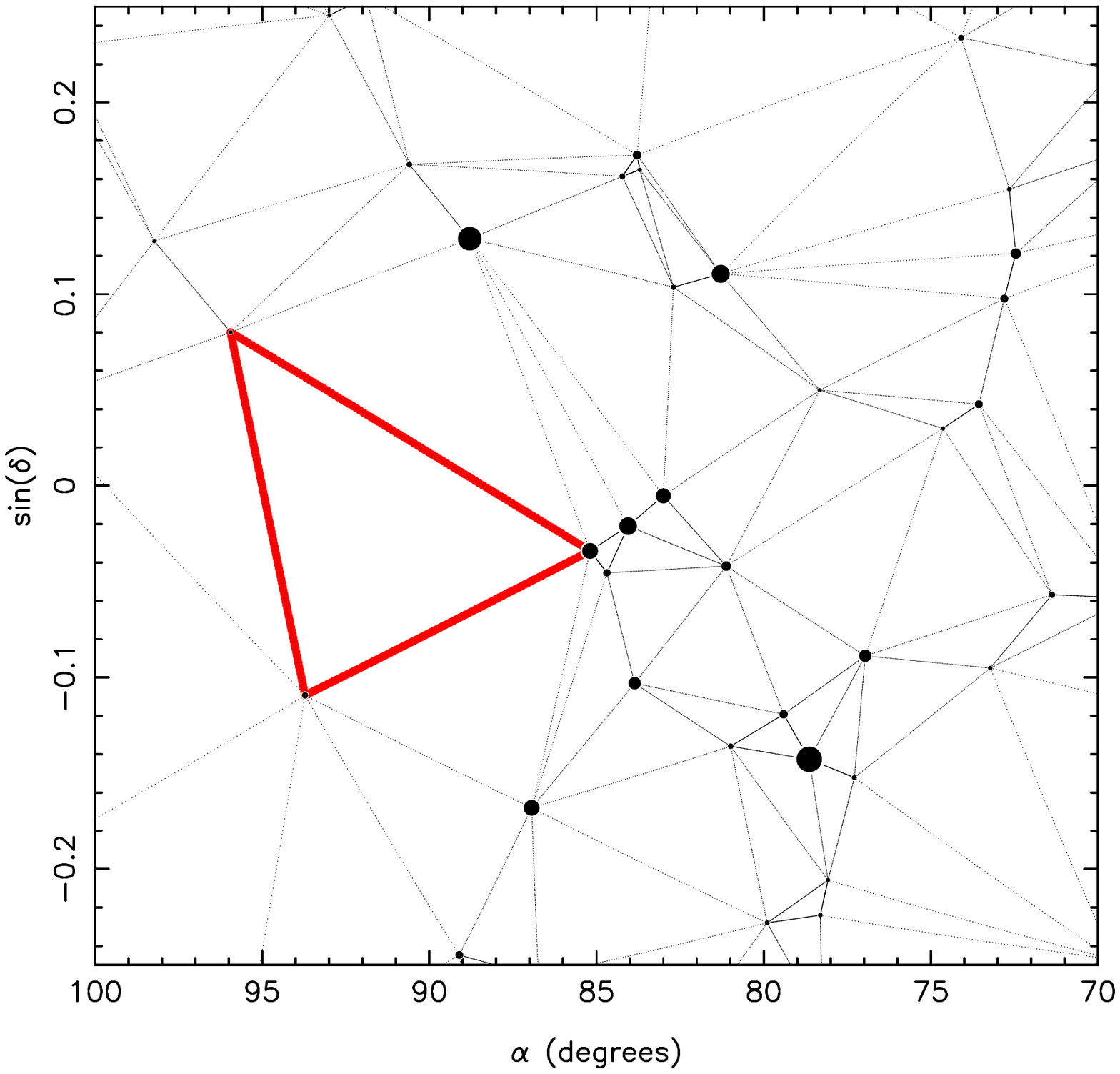}
\hfill
\includegraphics[viewport=150 78 659 566,width=\columnwidth]{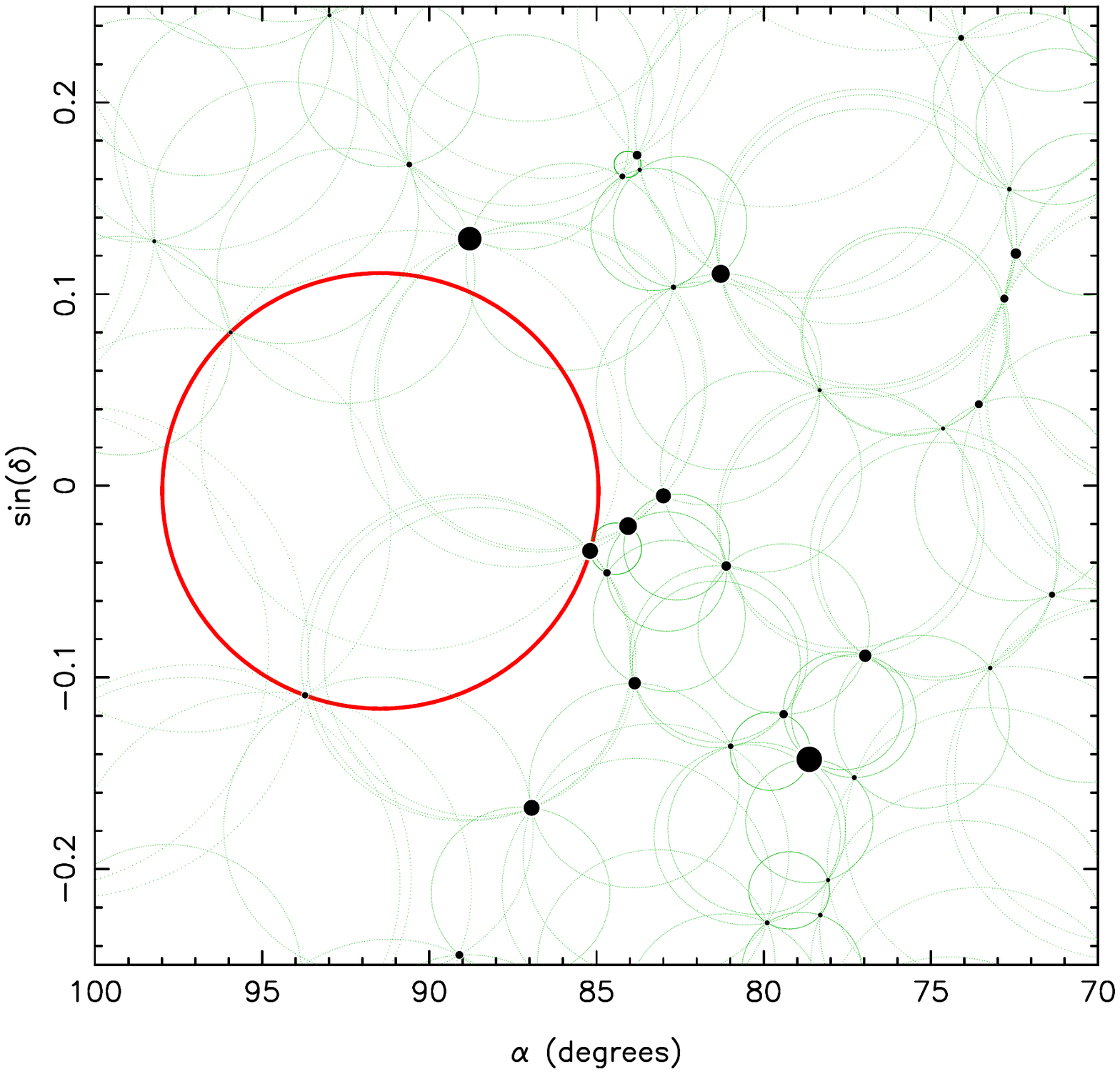}
\vskip 3mm
\caption{Application of the Delaunay triangulation to the celestial sphere.
\emph{Left:\/} Sky region centered at the Orion constellation
(\mbox{$\alpha=5^{\mbox{\scriptsize h}} 40^{\mbox{\scriptsize m}}$} and
\mbox{$\delta=0^{\circ}$}, J2000.0), represented using a Lambert's equal area
projection. All the stars brighter than $m_V=4.5$~mag have been
considered. Although each star is represented with filled circles of size
proportional to its brightness, the triangulation is carried out considering
all the stars as nodes for the triangles, independently of their magnitude. A
particular triangle is highlighted in red colour.
\emph{Right:\/}. Circumcircles corresponding to all the triangles derived in
the previous triangulation. The red circle indicates the circumcircle
associated to the previously highlighted triangle. Note that the empty
circumcircle interior property guarantees that there are no stars inside any of
the displayed circumcircles.}
\label{ini-triangulation}
\end{figure*}

The empty circumcircle interior property of the Delaunay triangulation
provides a straightforward method for a systematic search of regions
in the celestial sphere free from stars. If one computes the Delaunay
triangulation in the 2-dimensional surface of a sphere, using as nodes
the location of the stars down to a given threshold visual magnitude
($m$), the above property guarantees that all the circumcircles are
void of stars brighter than that magnitude (see
Fig.~\ref{ini-triangulation}). Thus, the circumdiameter
of every circumcircle determines the maximum field of view that can be
employed in that region of the sky as blank field.

In this work we have applied the Delaunay triangulation to the
\mbox{Tycho-2} stellar catalogue. \mbox{Tycho-2} contains astrometric and
photometric information for the 2.5~million brightest stars in the
sky, and it is complete up to magnitude {\it
  V}~=~11.5~mag. Photometric data consists of two pass-bands ({\it
  B$_T$} and {\it V$_T$}, close to Johnson B and~V;
\citealt{Perryman97a}).  Typical uncertainties are 60~mas in position
and 0.1~mag in photometry. In addition to the main catalogue, we have
used the first \mbox{Tycho-2} supplement, which lists another
17,588~bright stars from the Hipparcos and \mbox{Tycho-1} Catalogues
which are not in \mbox{Tycho-2}.

In order to proceed with the triangulation, we have made use of
STRIPACK \citep{renka97}, a Fortran 77 software package that employs
an incremental algorithm to build a Delaunay triangulation of a set of
points on the surface of the unit sphere. For $N$ nodes, the storage
requirement for the triangulation is $13N$ integer storage locations
in addition to $3N$ nodal coordinates. The computation scales as
\mbox{$O(N \log N)$}. It is important
to highlight that the original software was written using
single-precision floating arithmetic, which turns out to be
insufficient when dealing with astronomical coordinates with accuracy
$\sim$~1~arcsec. For the work presented here, we have modified the
software code in order to use double-precision floating arithmetic,
which guarantees the proper computation of the triangulation when
working with star separations approaching a few arcseconds.

\subsubsection{Tessellating a collection of smaller subregions}
\label{smaller-subregions}

\begin{figure}
\centering
\includegraphics[width=\columnwidth]{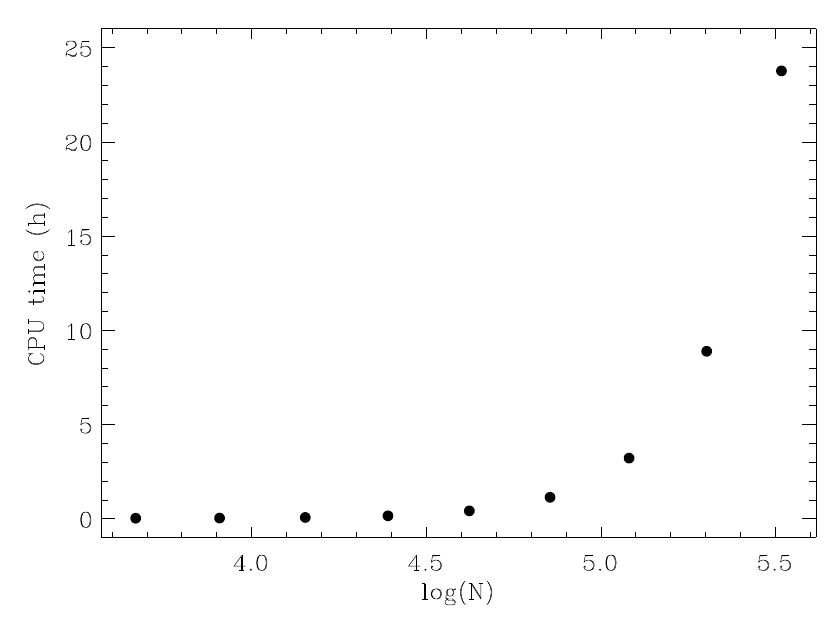}
\caption{CPU time consumed in function of the number of stars (nodes) used in
the tessellation. Each point corresponds to an increasing $m_{V_{T}}$ from 6
to~10 in steps of 0.5~mag.}
\label{CPUtime}
\end{figure}

In principle it is straightforward to obtain a list of blank fields for the
whole celestial sphere by using as input a stellar catalogue including stars
down to a given magnitude. However, considering that the number of stars grows
very rapidly with increasing limiting magnitude, this approach may turn out to
become nonviable, either in terms of computer memory or considering computing
time. In Fig.~\ref{CPUtime} we show the time taken by our computer in function
of the number of stars (nodes) to carry out the tessellation of the
\mbox{Tycho-2} catalogue. Each dot corresponds to an increase of 0.5~mag in
$m_{V_{T}}$, in the range from 6.0 to 10.0~mag. Although the CPU time depends
on the characteristic of the machine used for the tessellation, the most
important aspect is the exponential increase of the required time with
$m_{V_{T}}$, mainly due to the exponential growth of the number of stars (see
Table~\ref{result}). In addition, since the tessellation of the whole celestial
sphere demands memory storage (for stellar coordinates plus auxiliary
variables) that scales with the number of stars, this approach requires
excessively large memory capacity for faint magnitudes.

\begin{figure}
\centering
\includegraphics[viewport=124 78 677 566,width=0.93\columnwidth]{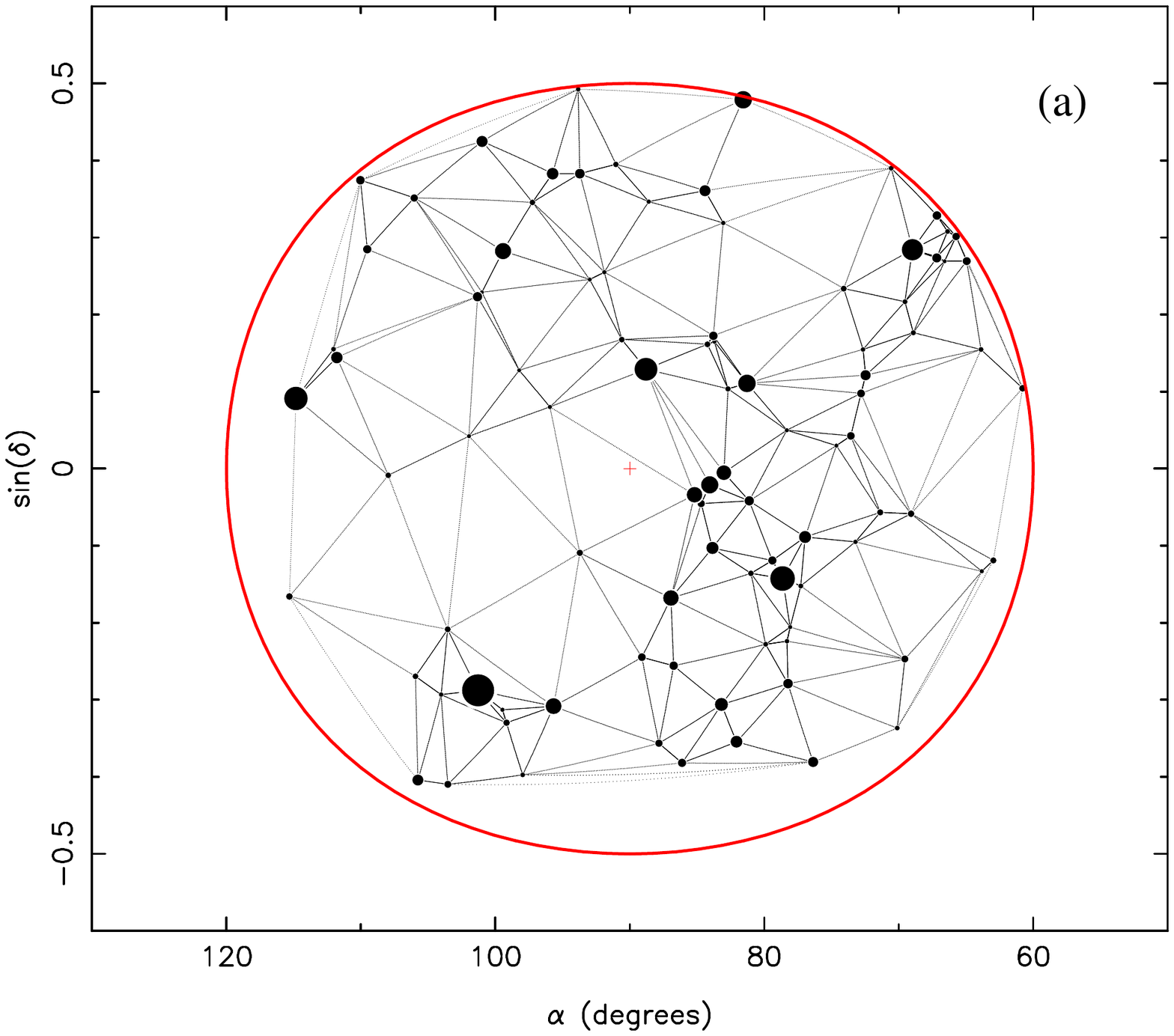}
\vskip 2mm
\includegraphics[viewport=124 78 677 566,width=0.93\columnwidth]{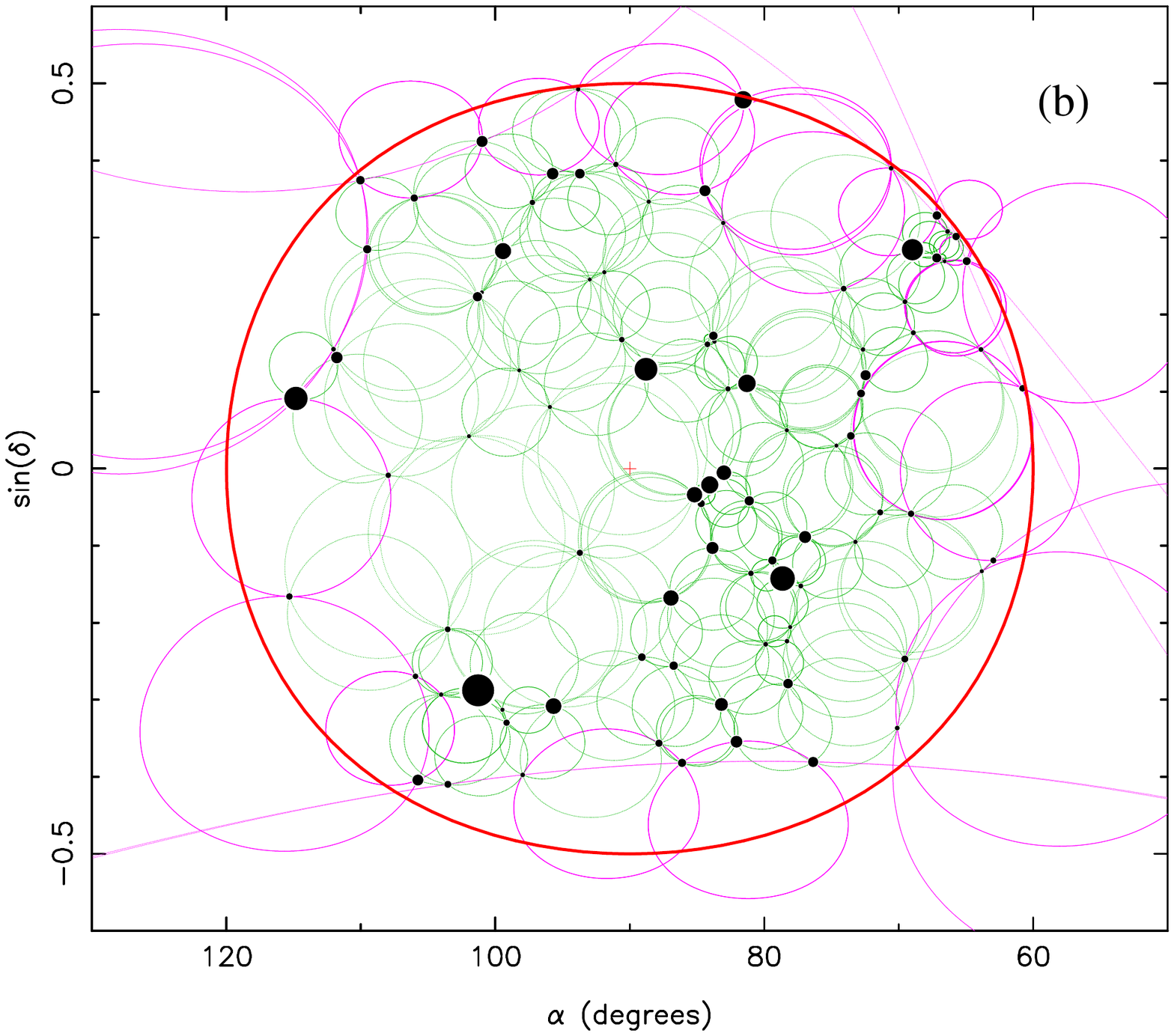}
\vskip 2mm
\includegraphics[viewport=124 52 677 566,width=0.93\columnwidth]{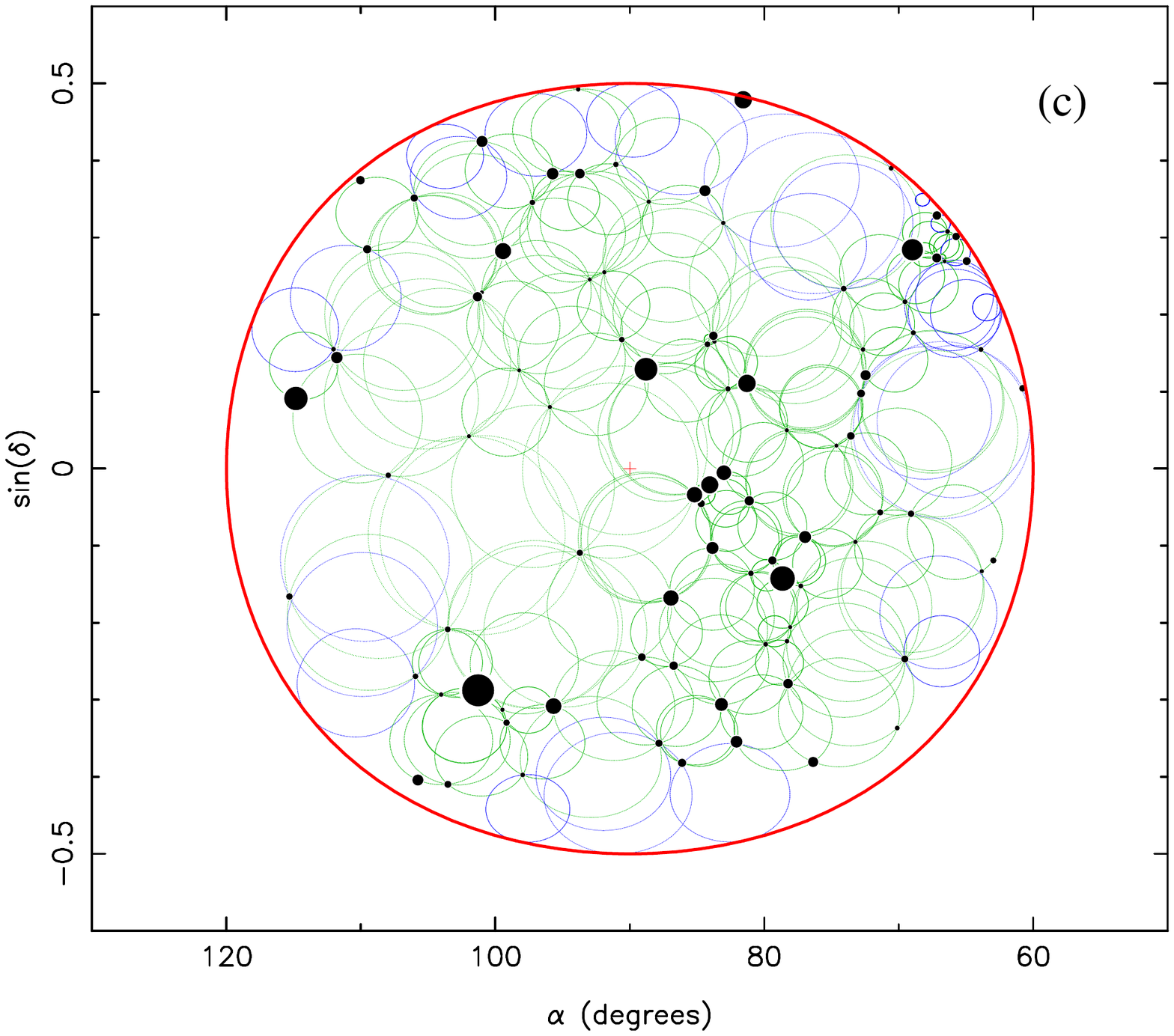}
\caption{Graphical illustration of the tessellation procedure when
  handling a subregion of the sky, in particular the portion of the
  celestial sphere (here represented using a Lambert's equal area
  projection) centered around the equatorial coordinates
  \mbox{$\alpha=6^{\mbox{\scriptsize h}}$} and
  \mbox{$\delta=0^{\circ}$} (J2000.0) within a radius of $30^{\circ}$,
  and using stars brighter than \mbox{$m_V=4.5$~mag}. See explanation
  in the text (section~\ref{smaller-subregions}).}
\label{figure:bt-problem}
\end{figure}

In these circumstances our approach is to subdivide the
whole celestial sphere in many smaller subregions (with some
overlap among them), in order to individually compute the Delaunay
triangulation in each of these smaller patches and, finally, merge the
separate blank field lists. However the merging process is not
immediate, since in this scenario the fact that the triangulation is
computed in small subregions have an additional side effect that must
be properly handled: when the subregion area is smaller than a single
hemisphere, the convex envelope of the stars within that subregion has
an interior zone (the subregion) and a larger exterior zone (the
celestial sphere excluding the considered subregion). This means that
the triangulation has boundary edges, i.e.\ triangles sides that are
not shared with other triangles \citep[see details in][]{renka97}. In
practice, the problem arises when one computes the circumcircles of
the boundary triangles, since in these cases those circumcircles may
easily expand beyond the limits of the considered subregion, where no
stars are present because they have not been included as input for the
triangulation. A graphical illustration of this problem is shown in
Fig.~\ref{figure:bt-problem}.

Figure~\ref{figure:bt-problem}a displays the initial triangulation
obtained after computing the Delaunay triangulation of the subregion
of the celestial sphere centered around the equatorial coordinates
\mbox{$\alpha=6^{\mbox{\scriptsize h}}$} and \mbox{$\delta=0^{\circ}$}
(J2000.0) within a radius of $30^{\circ}$, and using stars brighter
than \mbox{$m_V=4.5$~mag} (note that the Orion constellation is close
to the center of the field, whereas the bright stars Sirius, Procyon
and Aldebaran appear at the South-East, East and North-West,
respectively). The sky region is represented using a Lambert's equal
area projection \citep{2002A&A...395.1077C}. The thick red line
indicates the limit of the $30^{\circ}$~radius. Once the triangulation
has been computed, the next step is the calculation of the
circumcircle associated to each triangle.
Fig.~\ref{figure:bt-problem}b shows all the circumcircles obtained for
the triangulation derived in the previous step. The circumcircles that
are fully circumscribed within the $30^{\circ}$~radius are displayed
with green colour, whereas the circumcircles that expand beyond that
limit are represented in magenta.  It is obvious that the latter will
not be, in general, appropriate blank fields, since they have been
computed assuming that no stars were present beyond the red thick
line. However, even in these cases several possibilities can be
envisaged in order to obtain valid blank fields starting from the
information provided by the affected boundary triangles, while
respecting the constraint imposed by the proximity of the boundary of
the subregion. We discuss in
Appendix~\ref{appendix:boundary-triangles} several of these
possibilities. The result of following such approach is
displayed in Fig.~\ref{figure:bt-problem}c.

Finally, the blank field lists obtained for the different sky subregions
need to be merged. In our case, we have applied this procedure when
tessellating the sky for $m_{V_T}=10.5$ and~11.0~mag. In particular, we have
divided the whole celestial sphere in 578~circular subregions of~$8^{\circ}$
radius, where the centre of each subregion is separated by~$6^{\circ}$ in
declination and by~$6^{\circ}/\cos\delta$ in right ascension. Considering that
the maximum blank field radius obtained for $m_{V_T}=10.0$~mag is~$\sim
1^{\circ}$ (see Table~\ref{result}), this separation between subregions
provides an excellent overlap among them. Within the overlapping areas there
are duplicated blank fields and additional blank fields that were not permitted
to expand beyond their corresponding boundary limit. Consequently, to obtain
the final catalogue, we have removed both the repeated blank fields and those
due to the border effect (typically inscribed within larger blank fields of
neighbouring subregions). The latter were easy to identify because they
circumscribe less than 3 stars. Thus, the final all-sky blank field catalogue
is not affected by the number, size and overlap of the subregions that we have
employed.

\subsubsection{Preparing the nodes for the triangulation}

As described above, \mbox{Tycho-2} provides photometry data in two
pass-bands ({\it B$_T$} and {\it V$_T$}). We have applied the Delaunay
triangulation to each pass-band separately, and to the combination of
both. To obtain the initial collections of stars, we have selected
from both the main and the supplement \mbox{Tycho-2} catalogues all
stars with magnitude lower than $m$ in a given filter. In the
case of using both filters, the star magnitude should be lower than
$m$ in either of the two filters.

These initial collections of stars are still not suitable to compute
the Delaunay triangulation due to the presence of stars with
insufficient separation.
For that reason, we decided to
``merge'' into single objects all the stars closer than 1~arcsec,
similar to a typical seeing under good wheather conditions. The
resulting visual magnitude for the combined objects was computed as
the sum of the fluxes of the merged stars. The coordinates of the new
objects were placed in the line connecting the merged stars, closer to
the brightest star (using a weighting scheme dependent on the
individual brightness of the combined stars). This merging process
does not reduce substantially the final number of stars, but it makes
the triangulation process easier by removing unnecessary
very small triangles for which the computations are prone 
to rounding errors.

\subsection{Results}
\label{subsection:results}

We have applied the triangulation method to the merged star lists as
described above, with threshold magnitudes between 6 and~11~mag in steps of
0.5~mag. The resulting blank field catalogues are accessible through the
TESELA tool (explained in section~\ref{section:TESELA}).

\begin{figure*}
\centering
\includegraphics[viewport=40 106 744 562,width=0.32\textwidth]{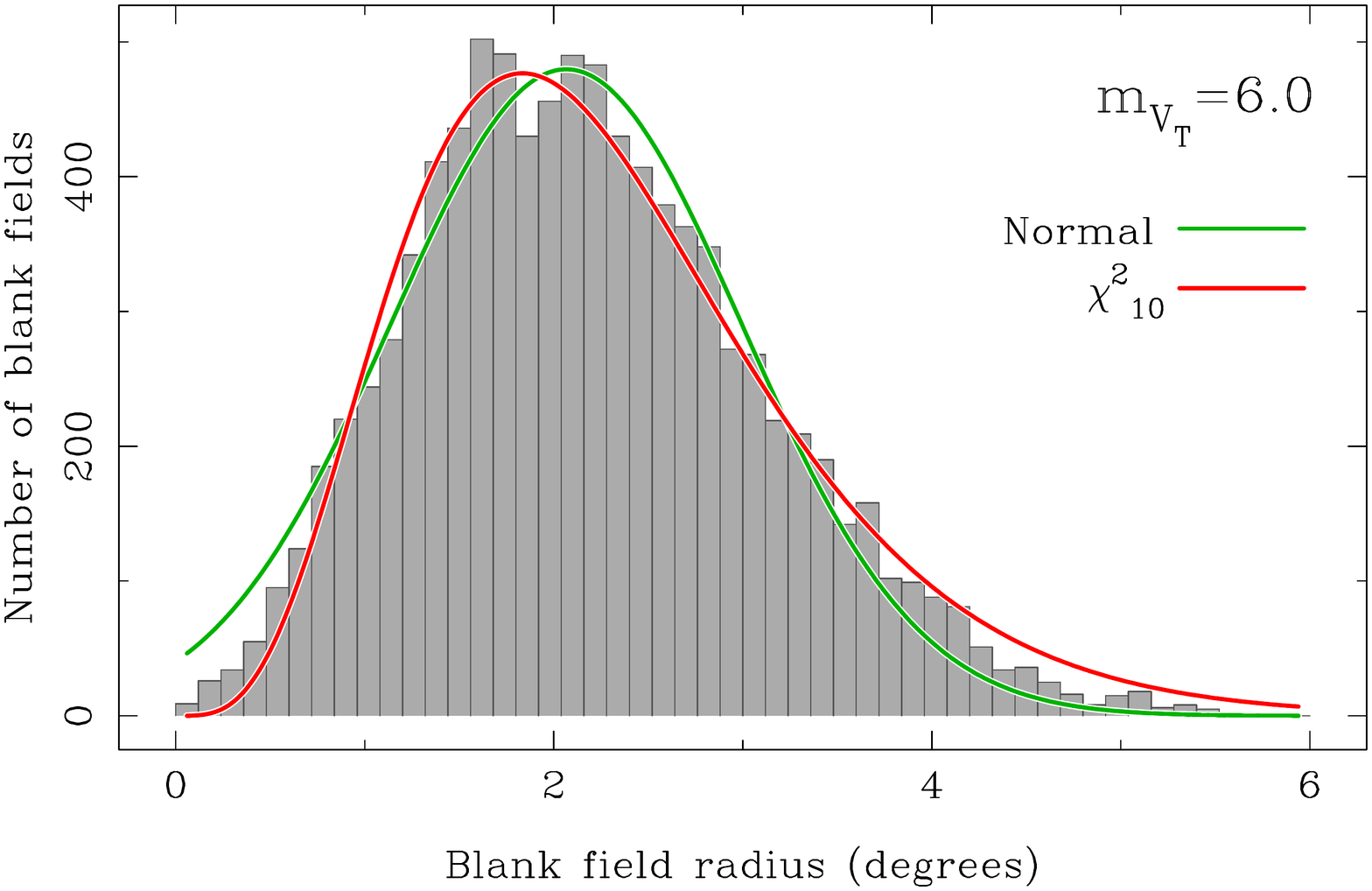}
\hfill
\includegraphics[viewport=40 106 744 562,width=0.32\textwidth]{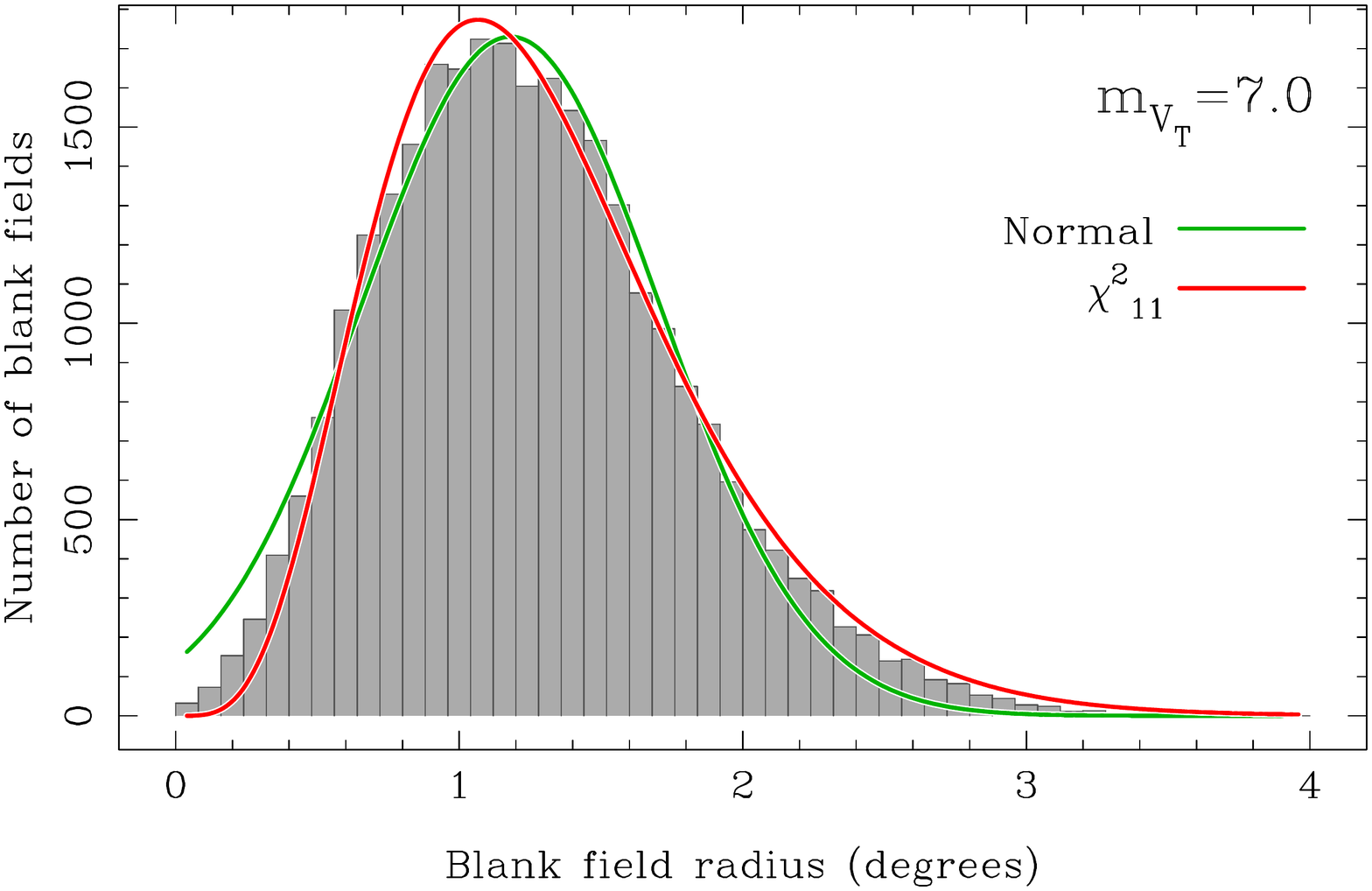}
\hfill
\includegraphics[viewport=40 106 744 562,width=0.32\textwidth]{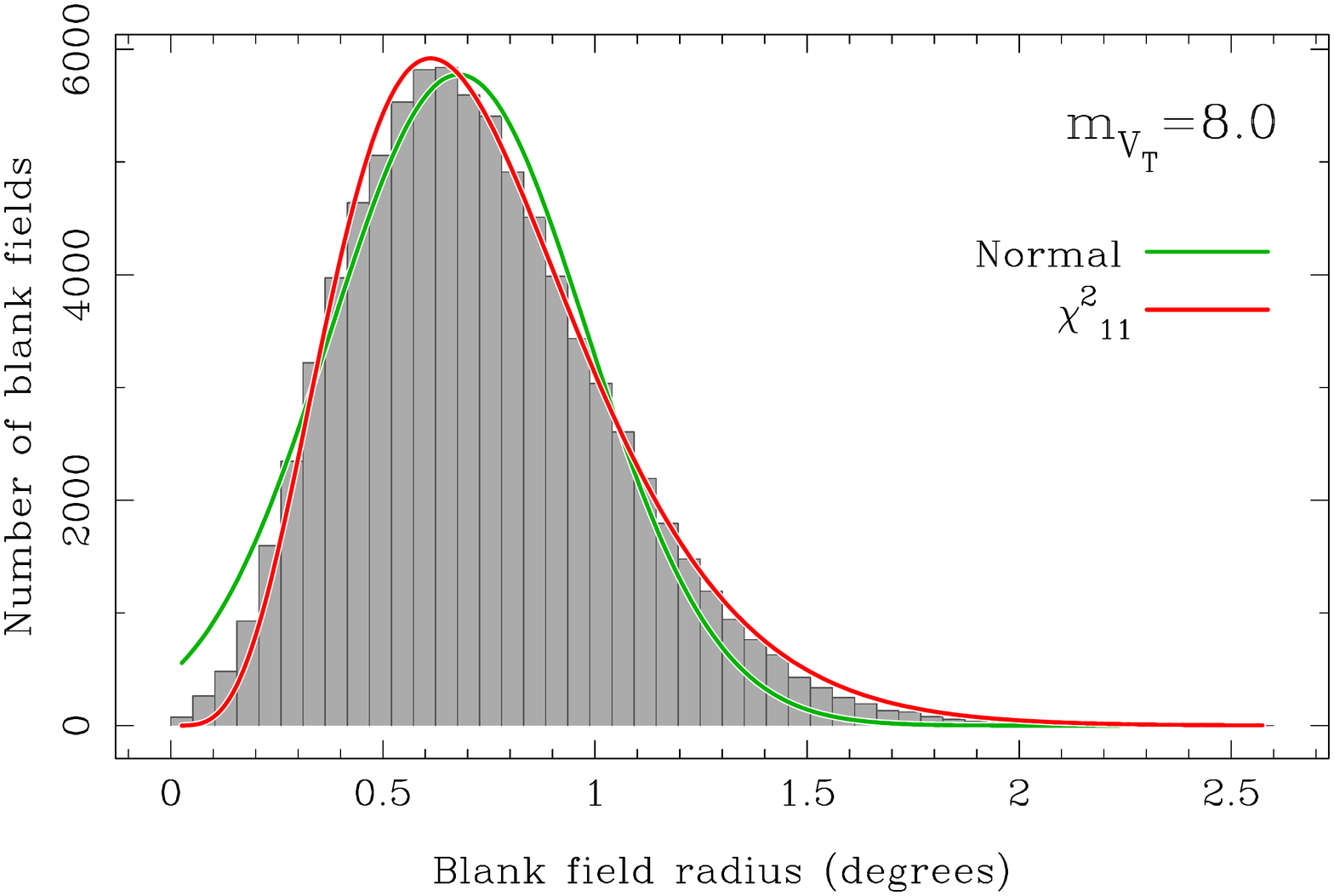}
\vskip 4mm
\includegraphics[viewport=40 106 744 562,width=0.32\textwidth]{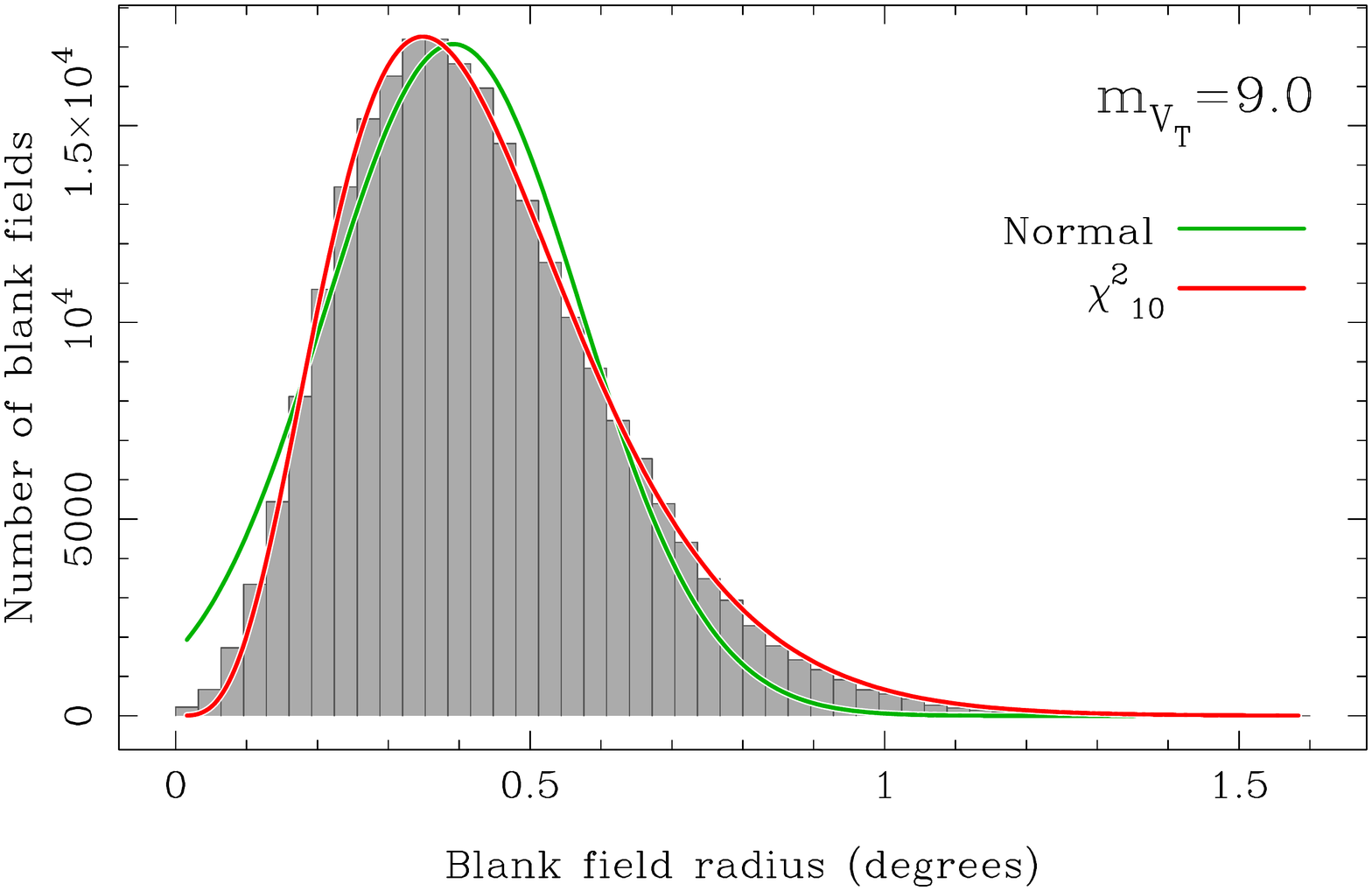}
\hfill
\includegraphics[viewport=40 106 744 562,width=0.32\textwidth]{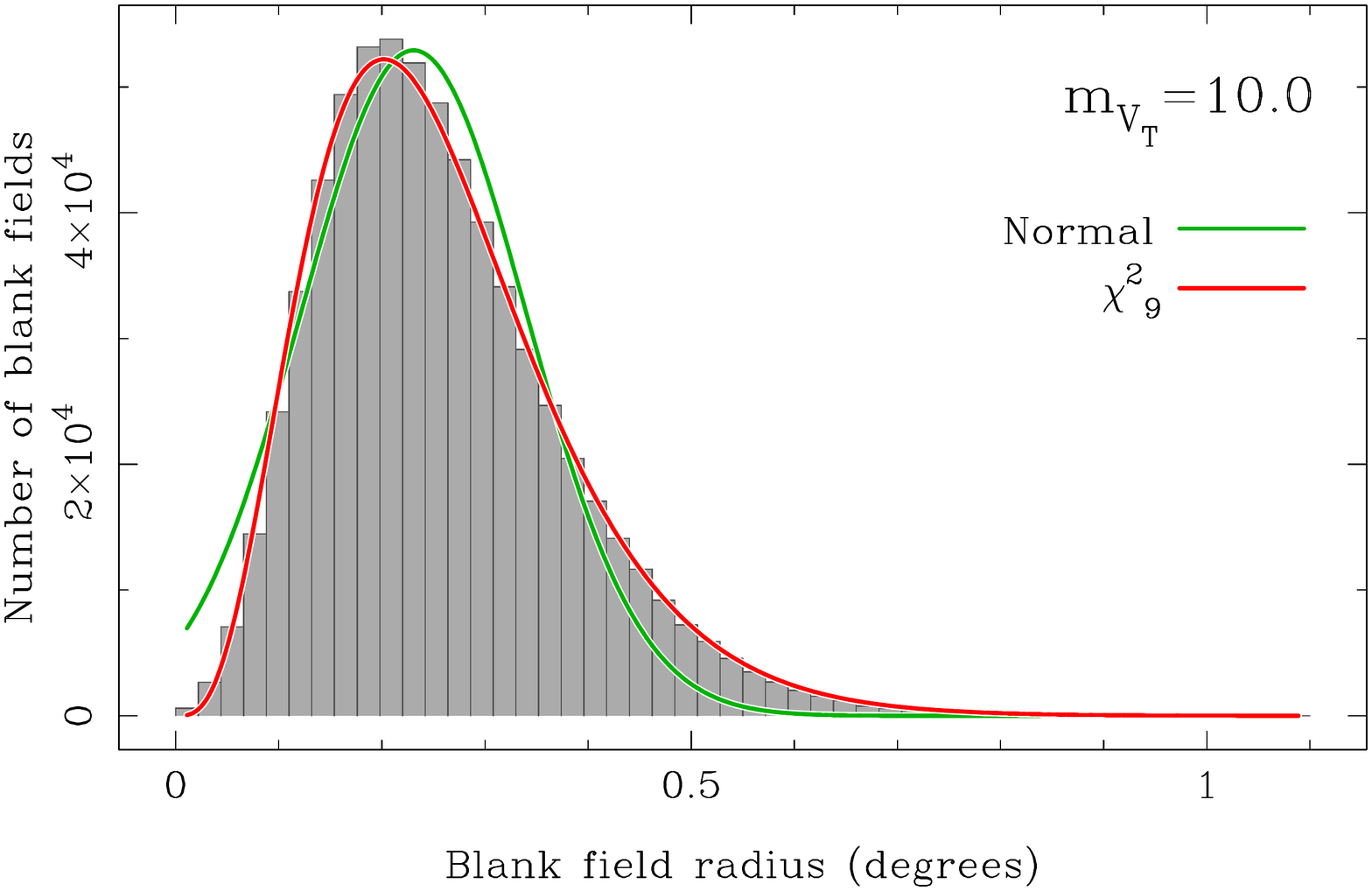}
\hfill
\includegraphics[viewport=40 106 744 562,width=0.32\textwidth]{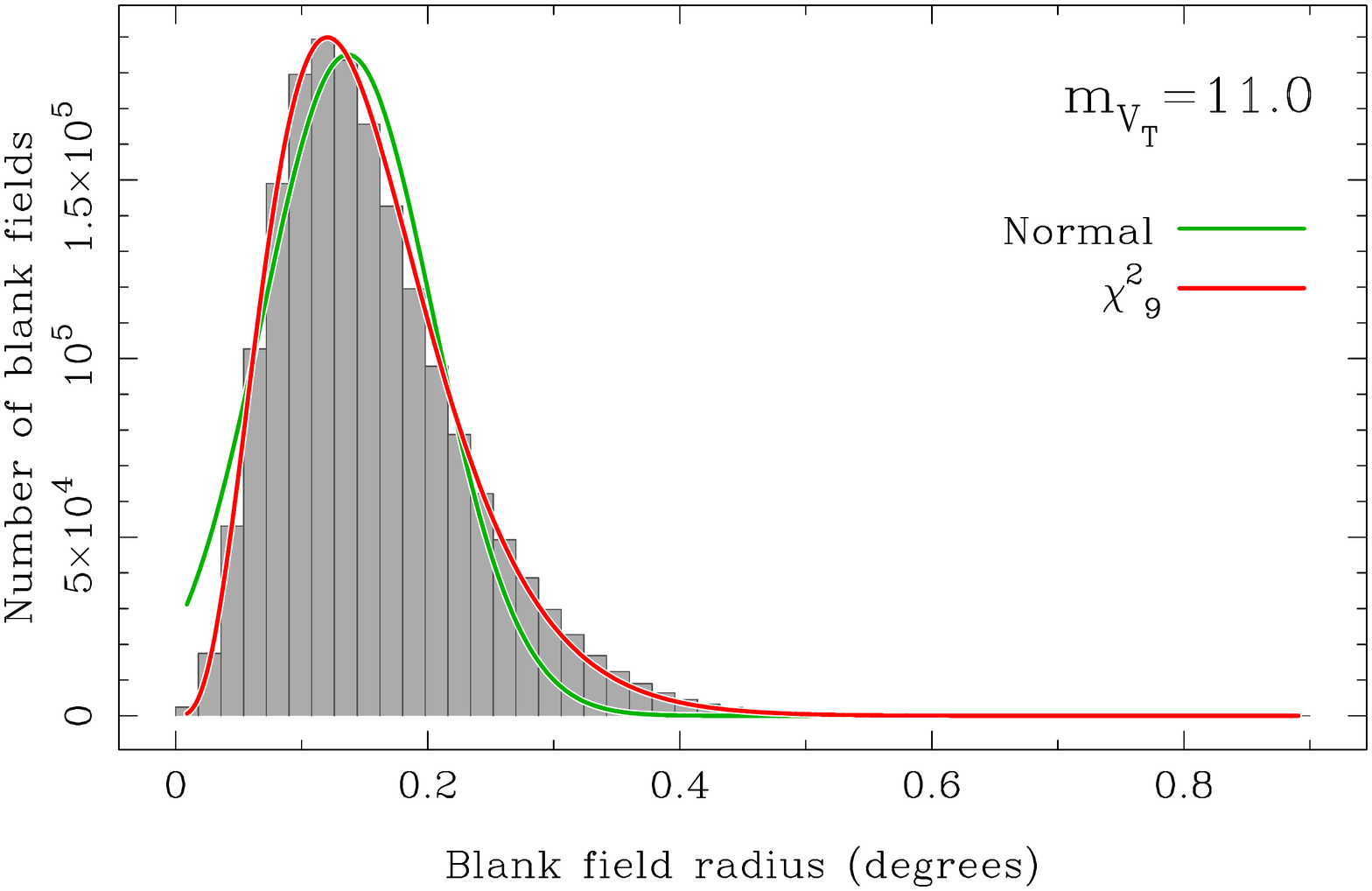}
\vskip 4mm
\caption{Histograms with the distributions of blank field radius, for different
values of $m_{V_T}$, as indicated in the upper right inset of each diagram.
Note that the x-axis range decreases as $m_{V_T}$ increases. The green curve is
the fit to a normal distribution, whereas the red line indicates the
corresponding fit to a $\chi^2_{\nu}$ distribution with
$\nu$~degrees of freedom, respectively. The coefficients of these parametric
fits are given in Table~\ref{result}. The best fit in all the diagrams is
obtained for the $\chi^2_{\nu}$ distribution.}
\label{figure:histograms}
\end{figure*}

\begin{table*}
  \caption[]{Results of the tessellation process using the
    \mbox{Tycho-2} catalogue with the {\it V$_T$} filter. Down to
    $m_{V_{T}}=10$~mag, the Delaunay triangulation has been obtained
    using the resulting stellar catalogues corresponding to the whole
    celestial sphere. For $m_{V_{T}}=10.5$ and~11.0, the results
    correspond to combining the individual triangulation of smaller
    sky subregions, as explained in the text (see
    section~\ref{smaller-subregions}). The last five rows of this
    table indicate the coefficients corresponding to the linear
    regressions $y=a+bx$, where $x=m_{V_{T}}$ in all the cases, and
    $y$ has been set to the logarithm (with base 10) of the parameters
    of the remaining columns where the linear fit makes sense. See
    description of the different columns in Sect.~\ref{subsection:results}.}
  \label{result}
  \begin{tabular}{rrrrrrrrrrrrr}
    \hline
    \hline
     \ccol{(1)}         & \ccol{(2)}     & \ccol{(3)}         & (4)               & (5)        & (6)                 & (7)               & (8)             & (9)              & (10)                & (11)             & (12)  &  \ccol{(13)}     \\
    \noalign{\smallskip}
     \ccol{$m_{V_{T}}$} & \ccol{$N_{*}$} & \ccol{$N_\dd{BF}$} & $\rho\dd{median}$ & $\Delta h$ & $A$                 & $\overline{\rho}$ & $\sigma_{\rho}$ &  $\rho_\dd{max}$ &                 $B$ &              $f$ & $\nu$ &  \ccol{CPU time} \\
     \ccol{(mag)}       &                &                    & (deg)             & (deg)      &                     & (deg)             &  (deg)          &   (deg)          &                     &                  &       &  \ccol{(h)}      \\
    \noalign{\smallskip} \hline \noalign{\smallskip}
                  6.0   &          4,648 &              9,292 &             2.111 &      0.120 & $4.796 \times 10^2$ &             2.067 &           0.928 &            5.805 & $4.879 \times 10^2$ &            0.436 &  10   &             0.02 \\
                  6.5   &          8,083 &             16,162 &             1.592 &      0.110 & $1.026 \times 10^3$ &             1.559 &           0.691 &            5.047 & $1.033 \times 10^3$ &            0.581 &  11   &             0.03 \\
                  7.0   &         14,229 &             28,454 &             1.211 &      0.080 & $1.730 \times 10^3$ &             1.182 &           0.525 &            3.702 & $1.747 \times 10^3$ &            0.767 &  11   &             0.06 \\
                  7.5   &         24,551 &             49,098 &             0.914 &      0.060 & $2.938 \times 10^3$ &             0.893 &           0.400 &            2.932 & $2.981 \times 10^3$ &            1.011 &  10   &             0.15 \\
                  8.0   &         41,989 &             83,974 &             0.696 &      0.052 & $5.776 \times 10^3$ &             0.679 &           0.302 &            2.510 & $5.830 \times 10^3$ &            1.335 &  11   &             0.41 \\
                  8.5   &         71,491 &            142,978 &             0.528 &      0.042 & $1.030 \times 10^4$ &             0.514 &           0.233 &            2.051 & $1.052 \times 10^4$ &            1.752 &  10   &             1.13 \\
                  9.0   &        120,381 &            240,758 &             0.405 &      0.032 & $1.707 \times 10^4$ &             0.392 &           0.180 &            1.557 & $1.767 \times 10^4$ &            2.294 &  10   &             3.21 \\
                  9.5   &        200,835 &            401,666 &             0.310 &      0.026 & $2.988 \times 10^4$ &             0.300 &           0.140 &            1.273 & $3.128 \times 10^4$ &            2.995 &  10   &             8.88 \\
                 10.0   &        328,819 &            657,634 &             0.240 &      0.022 & $5.291 \times 10^4$ &             0.231 &           0.109 &            1.051 & $5.570 \times 10^4$ &            3.850 &   9   &            23.76 \\
    \hline
                 10.5   &        538,719 &          1,077,434 &             0.185 &      0.020 & $1.005 \times 10^5$ &             0.177 &           0.086 &            0.943 & $1.077 \times 10^5$ &            4.998 &   9   &            ---   \\
                 11.0   &        871,336 &          1,742,668 &             0.143 &      0.018 & $1.850 \times 10^5$ &             0.137 &           0.068 &            0.853 & $2.026 \times 10^5$ &            6.460 &   9   &            ---   \\
    \hline
    \hline
            \ccol{$a$}  &         0.96   &             1.27   &            1.719  &       ---  &    \nullminus0.31   &            1.725  &          1.31   &           1.80   &    \nullminus0.35   & \nullminus1.753  &  ---  & \nullminus6.7    \\
     \ccol{$\Delta a$}  &         0.03   &             0.03   &            0.011  &       ---  &              0.04   &            0.009  &          0.02   &           0.05   &              0.04   &           0.012  &  ---  &           0.2    \\
            \ccol{$b$}  &         0.455  &             0.455  &  \nullminus0.2339 &       ---  &              0.506  &  \nullminus0.2362 &\nullminus0.227  & \nullminus0.176  &              0.513  &           0.2340 &  ---  &           0.80   \\
     \ccol{$\Delta b$}  &         0.004  &             0.004  &            0.0013 &       ---  &              0.004  &            0.0011 &          0.002  &           0.006  &              0.005  &           0.0014 &  ---  &           0.03   \\
            \ccol{$r^2$}&         0.9994 &             0.9994 &            0.9997 &       ---  &              0.9992 &            0.9998 &          0.9990 &           0.9901 &              0.9993 &           0.9997 &  ---  &           0.9909 \\
  \noalign{\smallskip}
  \hline
  \end{tabular}
\end{table*}

As an example, we present the results of the triangulation for {\it V$_T$} in
Fig.~\ref{figure:histograms}. The different histograms correspond to the
distributions of the number of blank fields as a function of the blank field
radius, for different threshold magnitudes $m_{V_T}$. The resulting histograms
are positively skewed and although, as a first order approach, they can
approximately be fitted with a normal distribution (green curves), better fits
are obtained using $\chi^2_{\nu}$ functions with $\nu$ degrees of freedom (red
curves). Note that the use of a chi-square law is just an empirical result,
obtained after trying to fit different well-known skewed distributions, like
lognormal and $F_{\nu_1,\nu_2}$ distributions (the latter with $\nu_1$ and
$\nu_2$ degrees of freedom), and finding that the best fits were obtained using
$F_{\nu_1,\nu_2}$ when $\nu_2 \rightarrow \infty$, which is equivalent to use a
$\chi^2_{\nu_1}$ distribution \citep[see e.g.][]{Press2007}. It is not the goal
of this work to provide a physical justification for this behavior.

The quantitative description of the above results are presented in
Table~\ref{result}. The threshold magnitude $m_{V_{T}}$ is given in the first
column. The number of stars $N_{*}$ (i.e. nodes) and the number of blank field
regions found $N_\dd{BF}$ are listed in the second and third columns,
respectively. The fourth column indicates the median blank field radius
$\rho\dd{median}$ and column~(5) is the bin width $\Delta h$ for the histogram
distributions (some of them displayed in Fig.~\ref{figure:histograms}) employed
to derive the parametric fits given in the next columns.  The amplitude $A$,
mean radius $\overline\rho$ and standard deviation $\sigma_{\rho}$, obtained
from the fit of the each histogram to a normal distribution of the form
\mbox{$A \exp[-(\rho-\overline\rho)^2/(2 \sigma^2_{\rho})]$} are listed in
columns~(6) to~(8), respectively. The maximum blank field radius appears in
column~(9). The coefficients of the fit to a chi-square distribution of the
form \mbox{$B \, \chi^2_{\nu}(\rho/f)$}, where $B$ is the amplitude, $f$ is a
scaling factor for $\rho$, and $\nu$ is the number of degrees of freedom, are
given in columns~(10) to~(12). Finally, column~(13) indicates the CPU time
taken for the triangulation when tessellating the whole celestial sphere at
once.

As we have shown in Fig.\ref{CPUtime}, the required CPU computing time
increases exponentially with $m$. For this reason, we have derived the
Delanuay triangulation for the whole celestial sphere at once just up
to magnitude 10, and in overlaping smaller regions (as explained in
section~\ref{smaller-subregions}) for higher $m$.

\subsection{Analysis}
\label{subsection:analysis}

We have analysed the results of the triangulation for {\it V$_T$}. For
{\it B$_T$} and the combination of both filters the results are very
similar.

Not surprisingly, there is an excellent correlation between most of the
parameters listed in Table~\ref{result} and $m_{V_T}$. The correlations can be
very well fitted using linear regressions of the form \mbox{$y=a+bx$}, being
$x=m_{V_T}$ and defining $y$ as the base-10 logarithm of the considered
parameter. The intercept $a$ and slope $b$ of the regressions (together with
their associated uncertainties $\Delta a$ and $\Delta b$) are given at the
bottom of each column. The final entry indicates the coefficient of
determination $r^2$, which in all the cases reveals the excellent correlation
between the fitted data.

\begin{figure}
\centering
\includegraphics[angle=-90,width=\columnwidth]{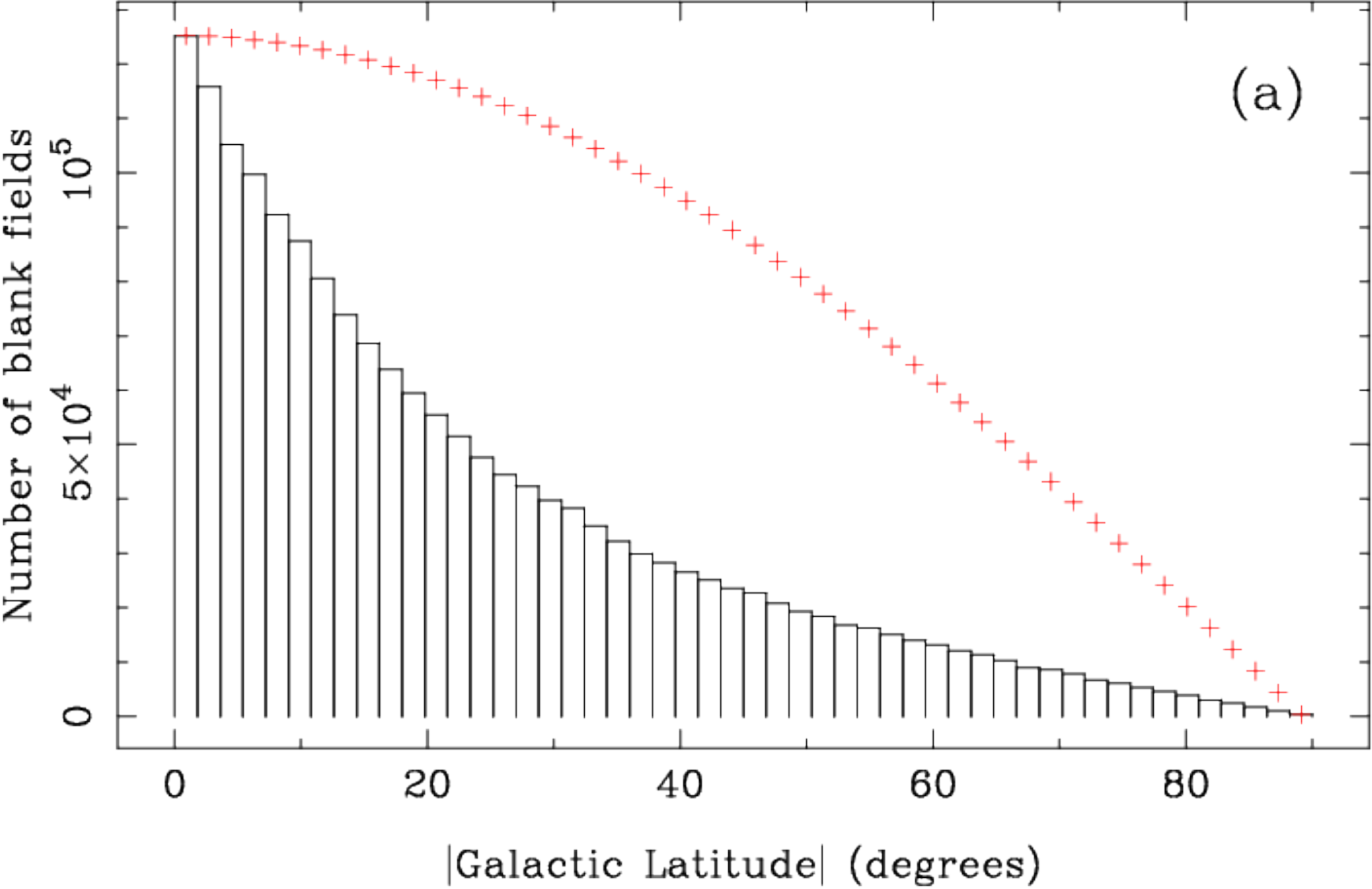}
\vskip 4mm
\includegraphics[angle=-90,width=\columnwidth]{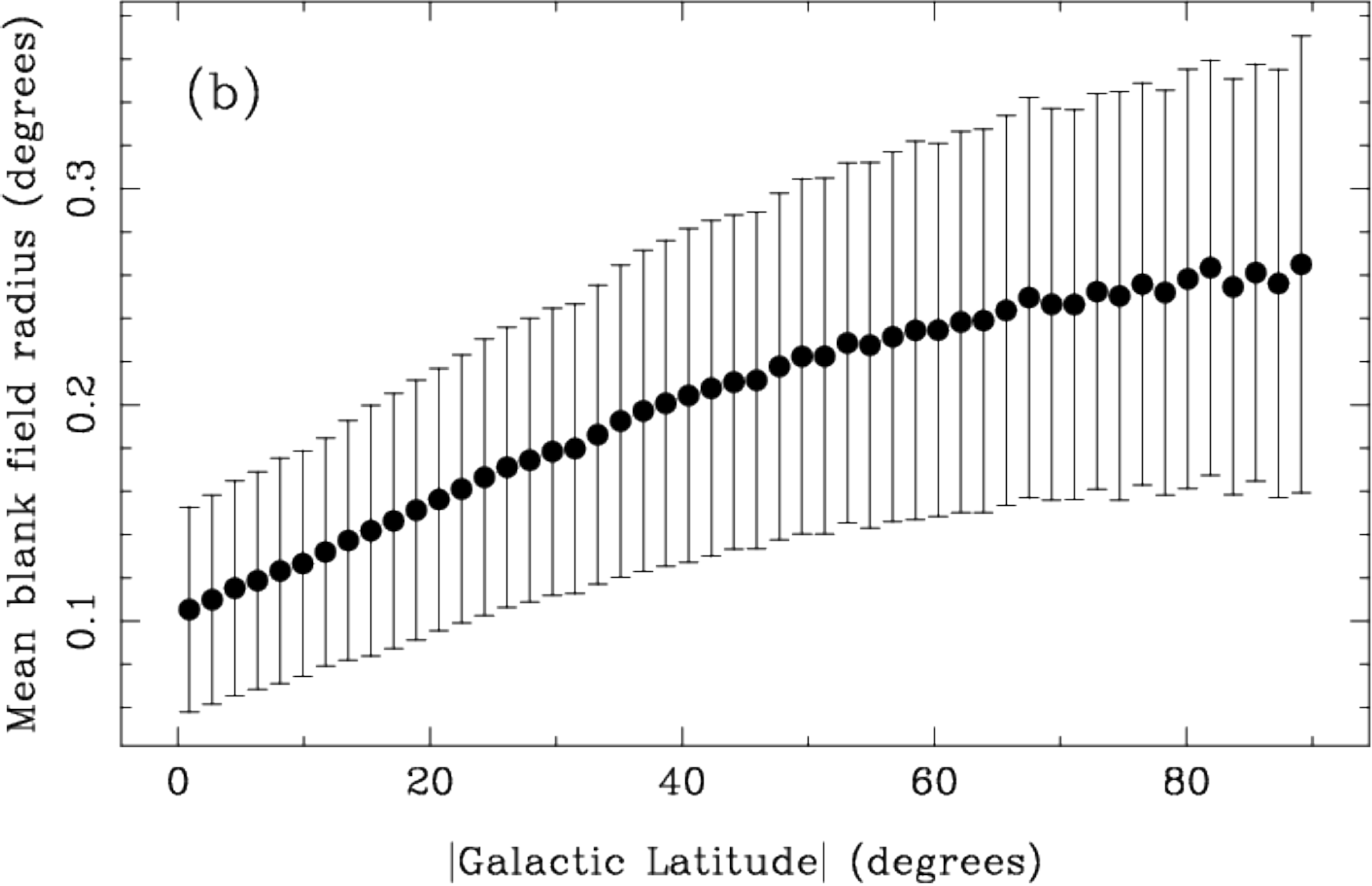}
\caption{Properties of the blank fields obtained in the case
\mbox{$m_{V_{T}}=11.0$~mag}.
{\bf Diagram~(a)}: number of blank fields as a function of the absolute value
of the galactic latitude. The red symbols show the relative variation of the
surface area on the celestial sphere with galactic latitude.
{\bf Diagram~(b)}: variation of the mean circumcircle radius as a function of
the absolute value of the galactic latitude (the error bars indicate the root
mean square exhibited by the radii at each galactic latitude bin).}
\label{figure:estadisticas}
\end{figure}

It is also interesting to examine the characteristics of the derived
blank fields at a given limiting magnitude. For example we have analyzed 
two illustrative diagrams corresponding to the case \mbox{$m_{V_{T}}=11.0$},
which are represented in Fig.~\ref{figure:estadisticas}.
Diagram.~\ref{figure:estadisticas}(a) displays the variation in the number of
regions as a function of the absolute value of galactic latitude $l$.  Not
surprisingly, the number of regions must decrease as the latitude increases,
since the area of a spherical annulus (shown with the red symbols) is maximum
at the galactic equator and tends to zero when $|l|$ approaches $90^{\circ}$.
However, this variation is not enough to explain the difference between the
histogram and the red symbols, which clearly indicates that the number of
regions is highly concentrated toward the galactic equator, which is simply the
result of the higher stellar densitiy in that region. The
diagram~\ref{figure:estadisticas}(b) shows that even though the number of
available blank fields unavoidably decreases with galactic latitude, their
typical size, and the variation of sizes among them, increase with $l$
approaching the galactic poles.

The extrapolation to fainter magnitudes of the exponential variations of
the parameters listed in Table~\ref{result}, together with the unavoidable
increasing difficulty when approaching low galactic latitudes, strongly
supports the need for an automatic tool that helps to identify suitable blank
fields when observing with medium/large size telescopes.

\section{TESELA}
\label{section:TESELA}

In order to provide an easy way to access the blank field catalogues, we
have created TESELA, a WEB accesible tool, developed within the
Spanish Virtual Observatory\footnote{\tt http://svo.cab.inta-csic.es/},
which can be publicly accessed through the following URL:\\ {\tt
http://sdc.cab.inta-csic.es/tesela.}

\begin{figure}
\centering
\fbox{\includegraphics[width=0.46\textwidth]{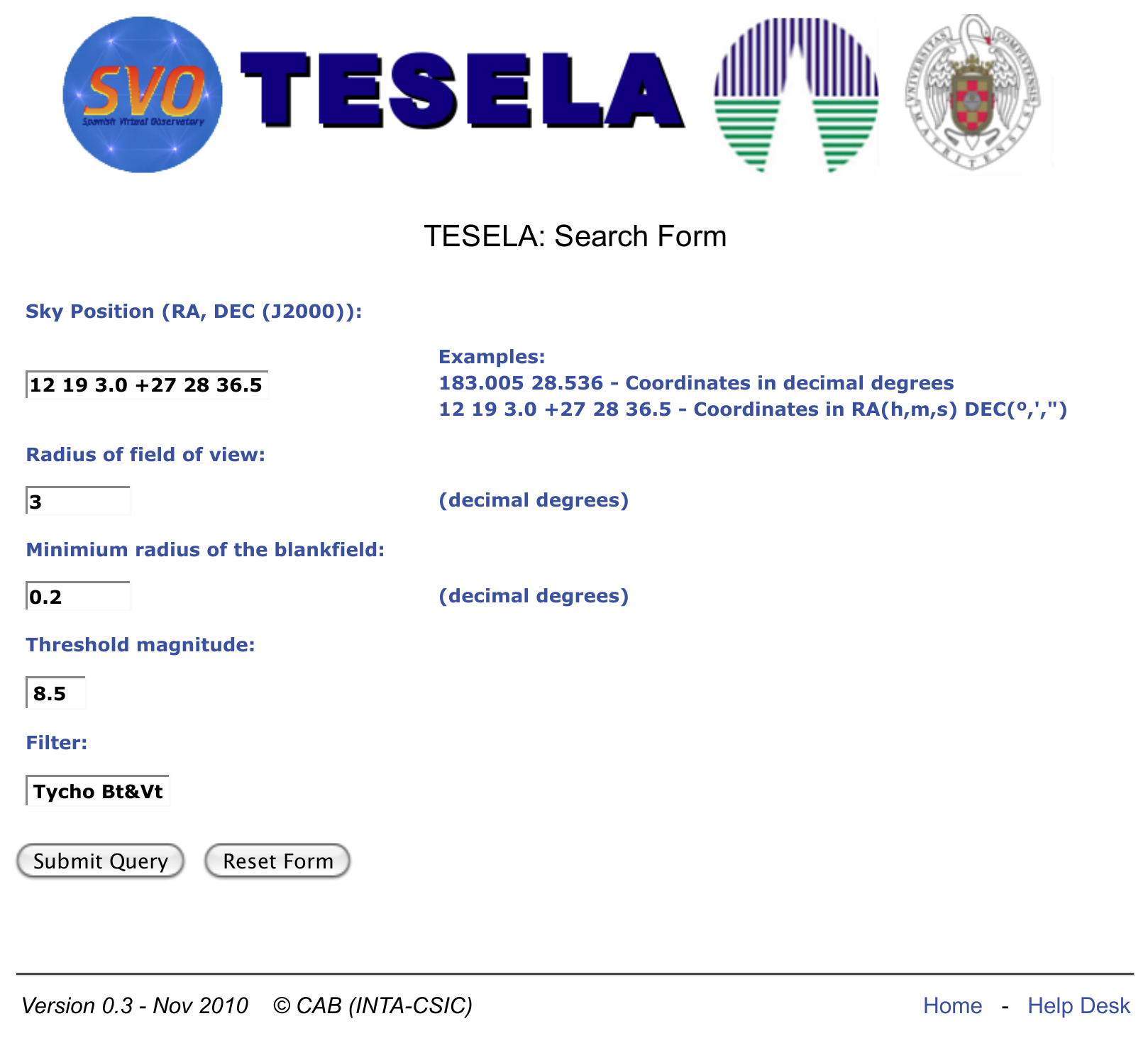}}
\caption{TESELA's search form. Through it the user defines the value
  of several parameters used for the cone-search. This form is available
  at\newline
  {\tt http://sdc.cab.inta-csic.es/tesela}}
\label{Tesela}
\end{figure}

The tool consists of a database containing the \mbox{Tycho-2}
stars and the already computed blank fields regions, and an
user-friendly interface for accesing the data. TESELA allows the users
to perform a cone-search around a position in the sky, 
i.e.\ to obtain the list of blank fields available around a given sky
position (right ascension and declination) and within a fixed radius around
that position.
Through its search form (see Fig.\ref{Tesela}), users can select the threshold
magnitude $m$, the \mbox{Tycho-2} filter to be used ({\it B$_T$}, {\it
  V$_T$}, or the combination of both), and define a minimum radius for
the blank field regions. This last option may be especially important
to fulfill observing requirements, e.g., to ensure that the blank
field is larger than the field of view of a given instrument.

%

TESELA presents the result of the cone-search in two tables, one for
blank fields (RA, DEC, radius) and the other for the \mbox{Tycho-2}
stars (RA, DEC, $m_{B_T}$ and/or $m_{V_{T}}$) in the searching
  area. These tables can be downloaded in CSV format for further
use. TESELA also provides users with the possibility of visualizing
the data. To do that, TESELA takes advantage of Aladin\footnote{\tt
  http://aladin.u-strasbg.fr/} \cite[][]{Bonnarel00}, a Virtual
Observatory (VO) compliant software that allows users to visualise and
analyse digitised astronomical images, and superimpose entries from
astronomical catalogues or databases available from the VO
services. Thanks to this connection with Aladin, we have provided
TESELA with the full capacity and power of the VO.

\begin{figure}
\centering
\includegraphics[width=0.47\textwidth]{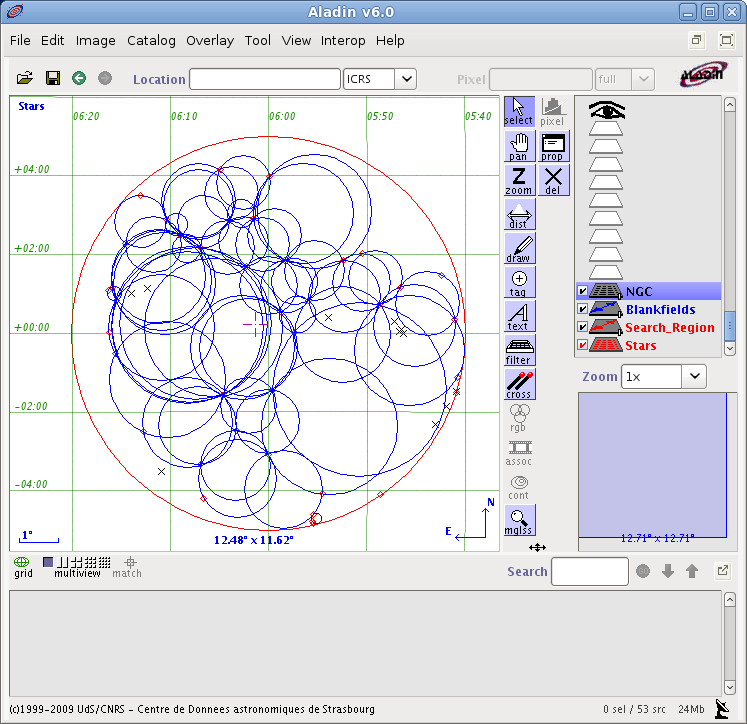}
\caption{Example of visualization of the data with Aladin. The large
  red circle defines the searching area, \mbox{Tycho-2} stars are
  shown by red diamonds, the blue circles depict the blank fields, and
  the black sails are the objects of the NGC~2000.0 in the
  region.}
\label{aladin}
\end{figure}
%

In Fig.~\ref{aladin} we show an example of how TESELA visualizes the
data using Aladin. TESELA sends to Aladin the \mbox{Tycho-2} stars of
the region, which are loaded in the first plane. The search area is
plotted in a second layer with a red circle; another layer depicting
the blank fields with blue circles is created, and finally, a last
layer shows the objects of NGC~2000.0 (The Complete New General
Catalogue and Index Catalogue of Nebulae and Star Clusters;
\citealt{Sinnott97}). Using this Aladin window, users are allowed to
load images and catalogues both locally and from the VO, having full
access to the whole universe resident in the VO. This can be very
helpful to determine the potential influence of relatively bright
nebulae and extragalactic sources in the sought regions. Note also
that, for obvious reasons, Solar System objects have been not
considered, and they must be taken into account in order to make use
of blank field regions close to the Ecliptic at a given date.

So far, the current version of TESELA allows users to access a
collection of blank fields obtained from the optical Tycho-2
catalogue. But Tesela has been conceived as a dynamic tool, which will
be improved in the future with both deeper optical catalogues and
catalogues in others wavelength ranges. Any future change in the tool
will be properly documented in its web site.

\section*{Acknowledgements}

We would like to thank the anonymous referee for the careful reading of
the manuscript and for her/his comments, which have helped to clarify this
paper. This work was partially funded by the Spanish MICINN under the
Consolider-Ingenio 2010 Program grant CSD2006-00070: First Science with the
GTC\footnote{\tt http://www.iac.es/consolider-ingenio-gtc}. This work was also
supported by the Spanish Programa Nacional de Astronom\'{\i}a y
Astrof\'{\i}sica under grants \mbox{AYA2008--02156} and \mbox{AYA2009-10368},
and by AstroMadrid\footnote{\tt http://www.astromadrid.es} under project
\mbox{CAM~S2009/ESP-1496}. This work has made use of Aladin developed at the
Centre de Donn\'ees Astronomiques de Strasbourg, France.

\appendix

\section{Handling boundary triangles}
\label{appendix:boundary-triangles}

Although tessellating small sky subregions is a good approach to the problem of
deriving blank fields when dealing with very large stellar catalogues, the
discussion of Section~\ref{smaller-subregions} revealed that the Delaunay
triangulation of such subregions leads to the situation displayed in
Fig.~\ref{figure:bt-problem}b, where many circumcircles associated to boundary
triangles (shown in magenta colour in that figure) clearly expand beyond the
limit of the considered subregion. Although in this case those circumcircles
can be easily discarded, it is possible to derive, starting from the
coordinates of the corresponding boundary triangles, new circles devoided of
stars that remain within the boundary of the sky subregion. This appendix
describes in more detail different approaches that can be employed in this
circumstance.

\begin{figure*}
\centering
\includegraphics[viewport=201 79 641 518,width=1.0\columnwidth]{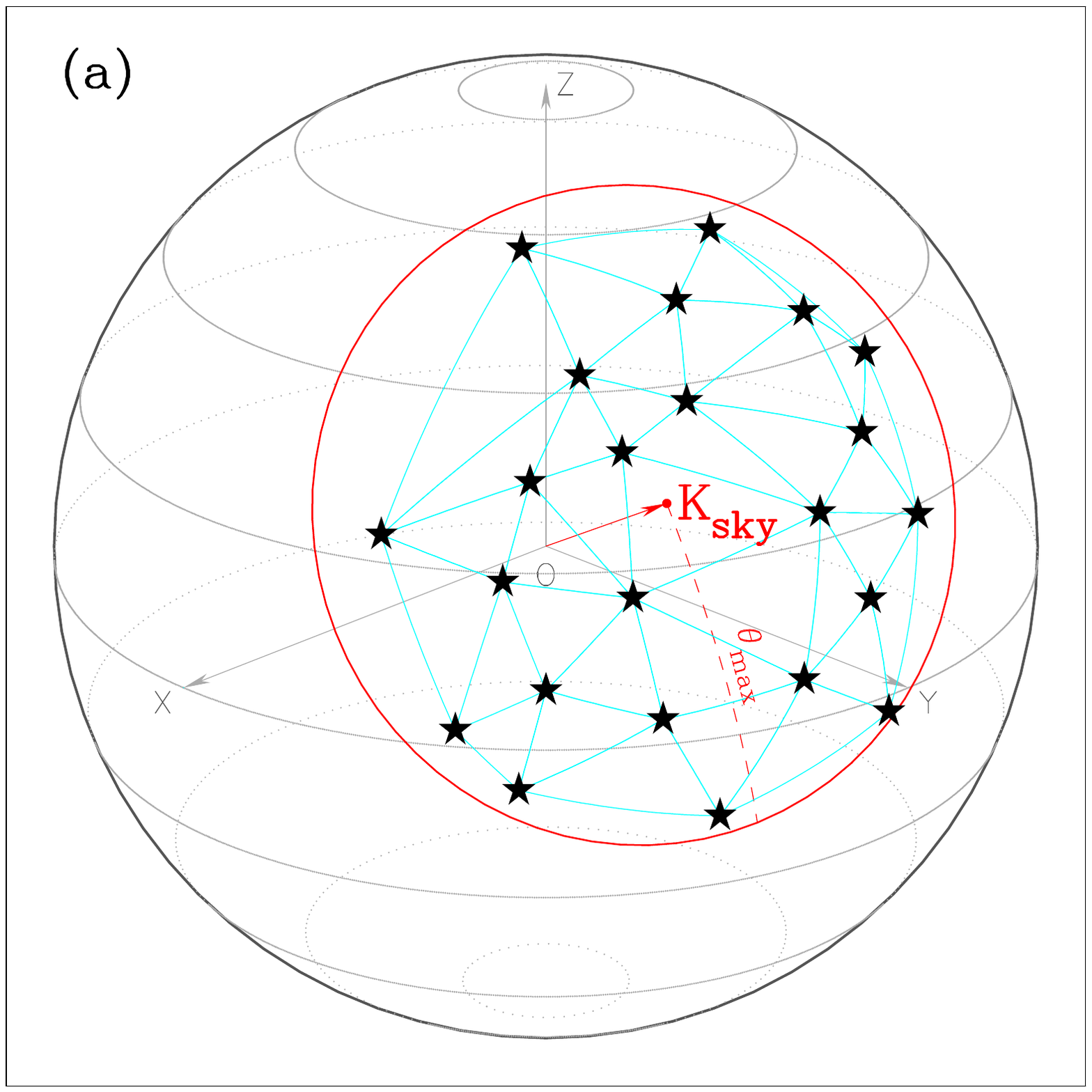}
\hfill
\includegraphics[viewport=201 79 641 518,width=1.0\columnwidth]{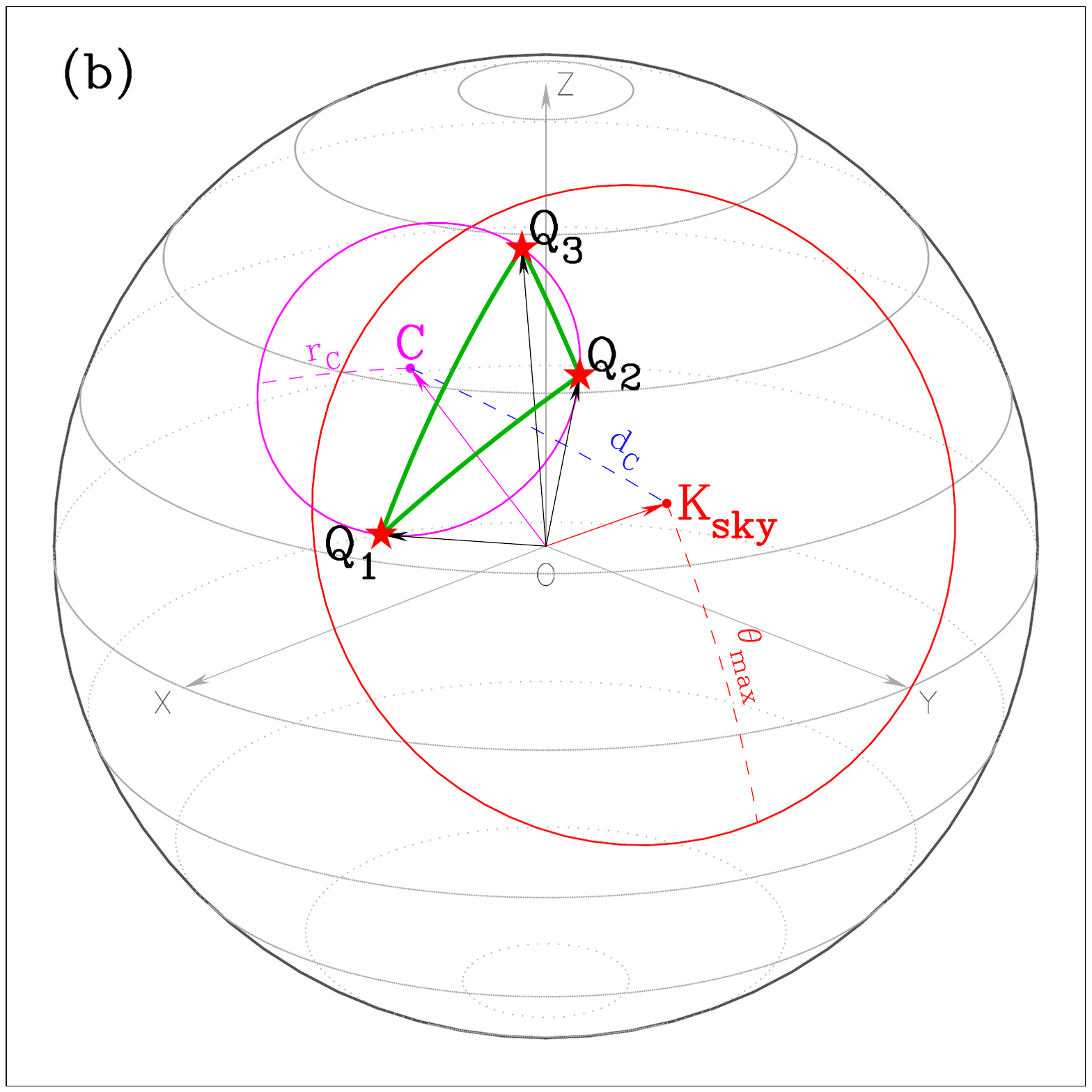}
\caption{Graphical illustration of the notation employed in
Appendix~\ref{appendix:boundary-triangles} to explain the initial steps when
handling boundary triangles. {\bf Diagram~(a):}~The whole celestial sphere of
unit radius and centered at the origin $\bmath{O}$ is displayed in gray. The
sky subregion under tessellation, which boundary is shown in red, is centered
at the point indicated by the unit vector $\bmath{K_\dd{sky}}$, and has a
radius $\theta_\dd{max}$ (dashed red line) measured along a great circle on the
surface of the celestial sphere. Only the stars contained within this region
are employed to compute the Delaunay triangulation (shown in cyan). {\bf
Diagram~(b):}~A~sample boundary triangle formed by the stars located at
$\bmath{Q_1}$, $\bmath{Q_2}$ and $\bmath{Q_3}$, given in counterclockwise
order, is represented in green. Its corresponding circumcircle, centered at
$\bmath{C}$ and passing through the three node stars, is displayed in magenta.
The radius of this circumcircle $r_C$ is plotted with a dashed magenta line.
The angular distance $d_C$ between $\bmath{C}$ and $\bmath{K_\dd{sky}}$ is
shown with a dashed blue line. In this example the circumcircle expands beyond
the limit of the sky subregion, i.e.\ the condition given in
Eq.~\ref{equation:condition-inside} does not hold.}
\label{figure:step0}
\end{figure*}

In what follows we are considering the celestial sphere as the unit sphere
centered at the origin $\bmath{O}$. The sky subregion where the blank fields
will be searched for is the surface of a spherical cap, which center is given
by the unit vector $\bmath{K_\dd{sky}}$, and the angular radius of such
subregion, as measured from $\bmath{O}$, is $\theta_\dd{max}$ (see
Fig.~\ref{figure:step0}). The Delaunay
triangulation will be computed with all the stars (down to a given magnitude)
within that sky subregion. Denoting the unit vectors poiting to these stars as
$\bmath{S_i}$ (with $i=1,\ldots,N_\dd{stars}$, being $N_\dd{stars}$ the total
number of stars within the subregion), it is obvious that the condition that
all these stars must satisfy is
\begin{equation}
\theta_\dd{max} \geq 
\mbox{arccos}(\bmath{K_\dd{sky}} \cdot \bmath{S_i}), 
\,\,\forall i=1,\ldots,N_\dd{stars}.
\end{equation}
We are also assuming that $\theta_\dd{max}< \pi/2$, i.e., the sky
subregion area under consideration is smaller than a single hemisphere.

\begin{figure*}
\centering
\includegraphics[viewport=201 79 641 518,width=0.66\columnwidth]{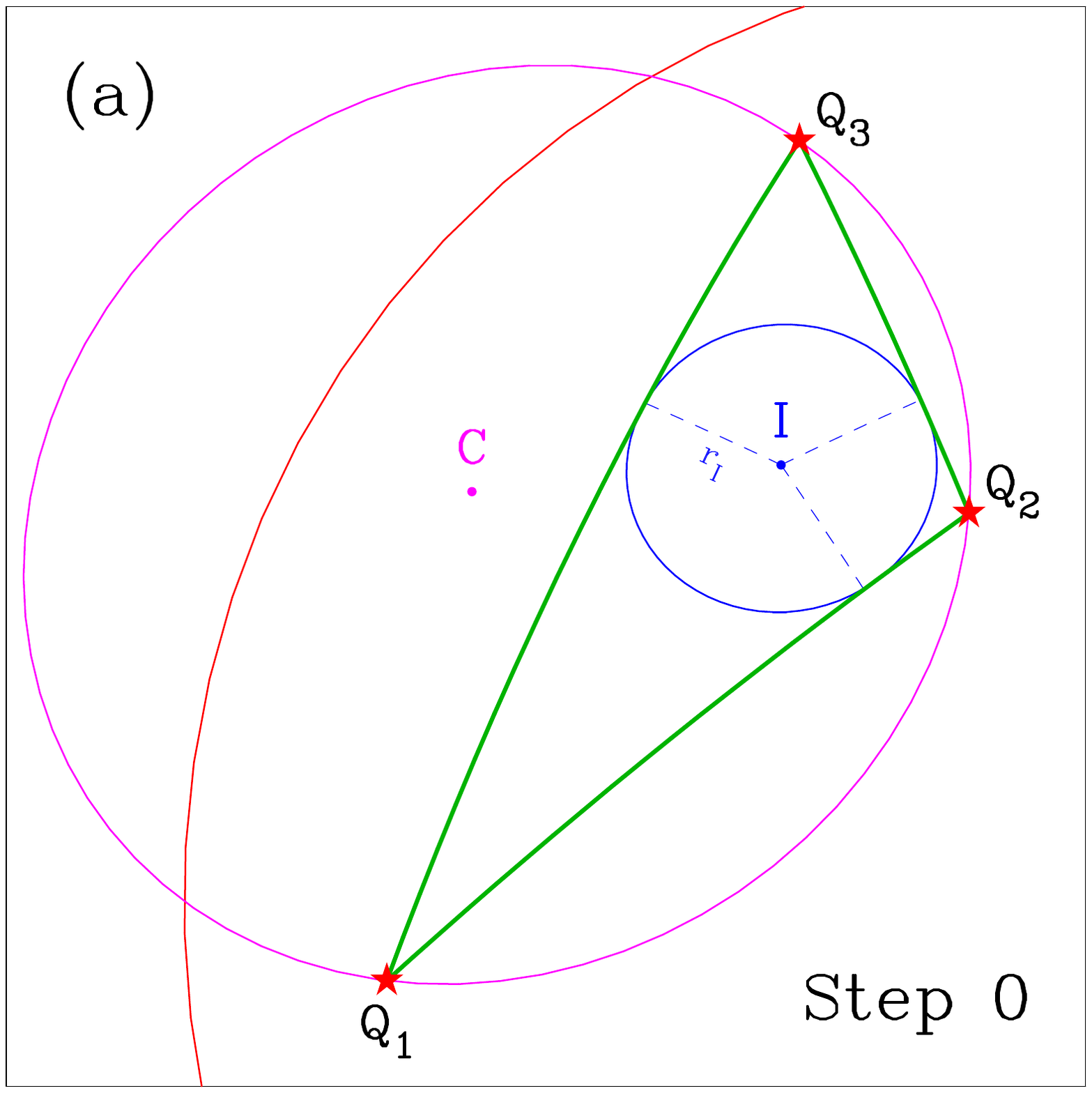}
\hfill
\includegraphics[viewport=201 79 641 518,width=0.66\columnwidth]{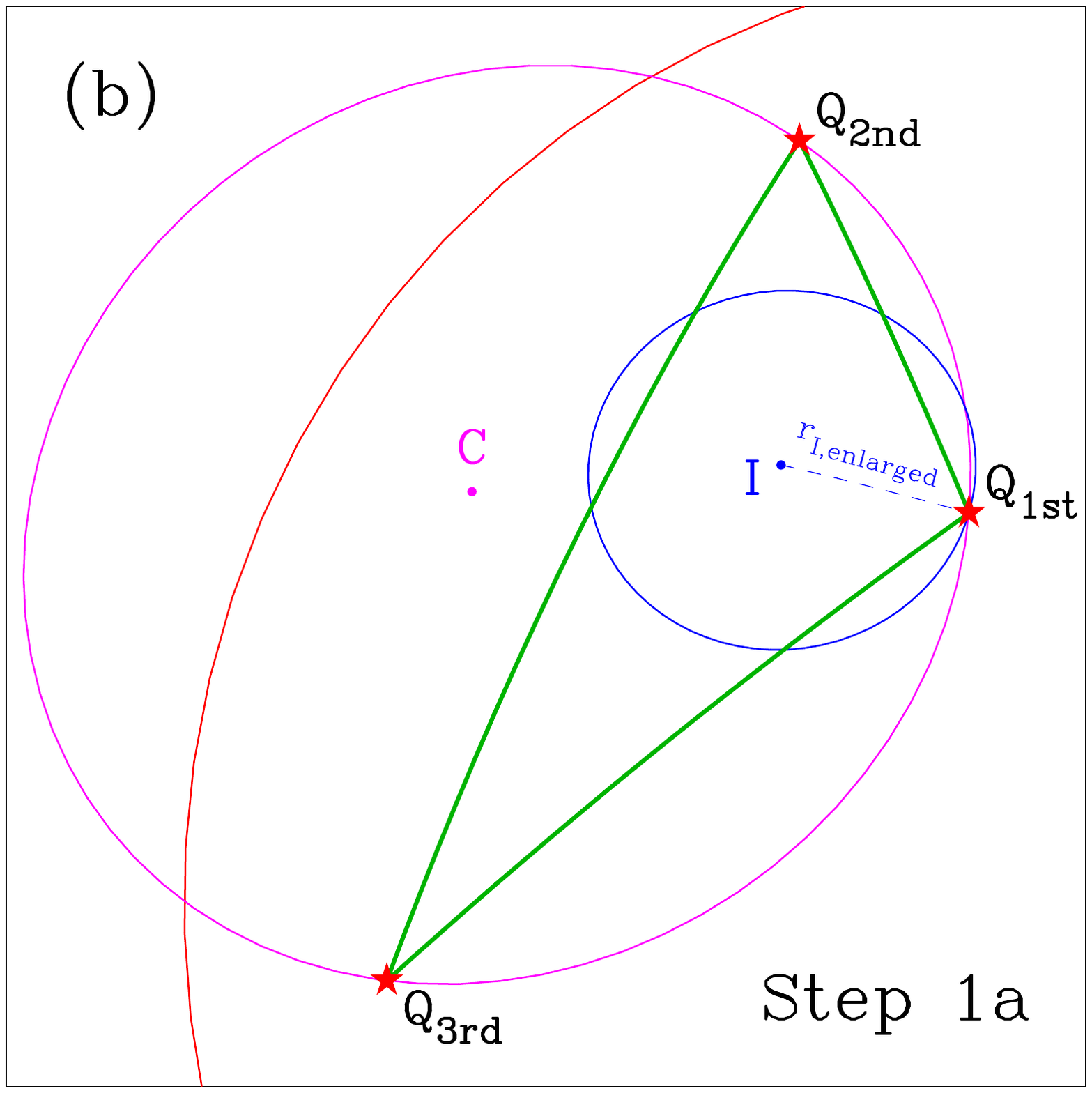}
\hfill
\includegraphics[viewport=201 79 641 518,width=0.66\columnwidth]{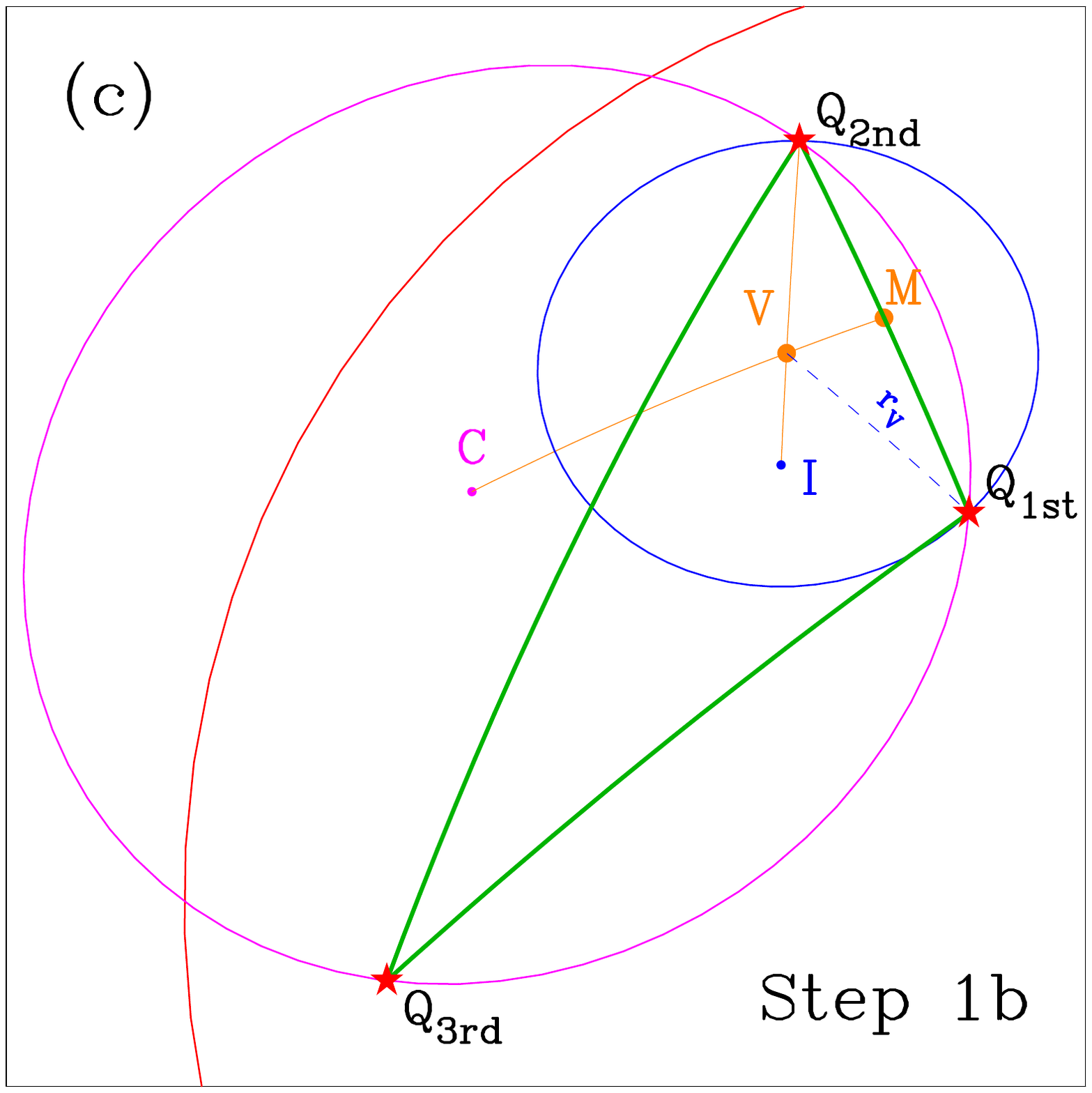}
\vskip 5mm
\includegraphics[viewport=201 79 641 518,width=0.66\columnwidth]{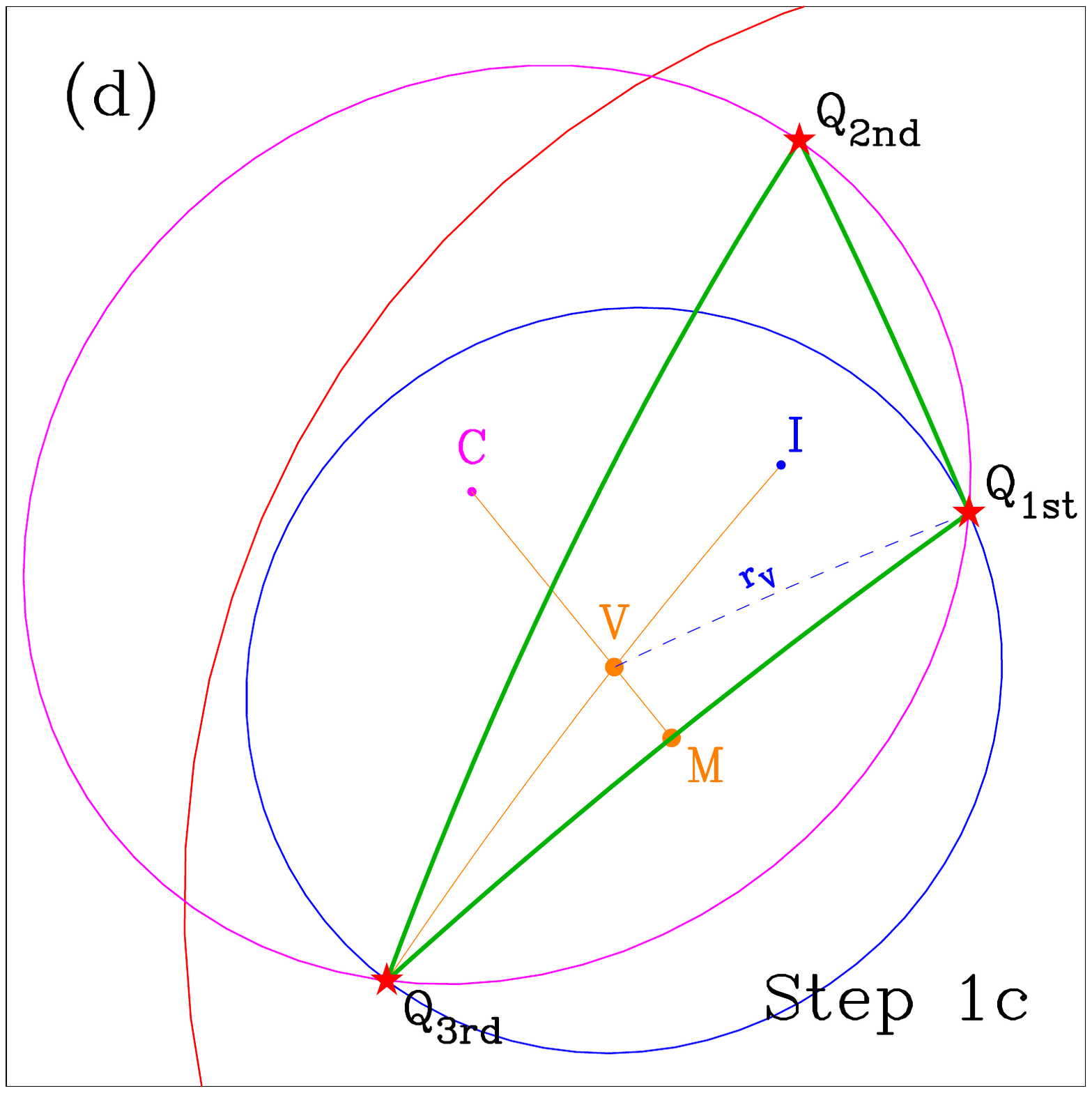}
\hfill
\includegraphics[viewport=201 79 641 518,width=0.66\columnwidth]{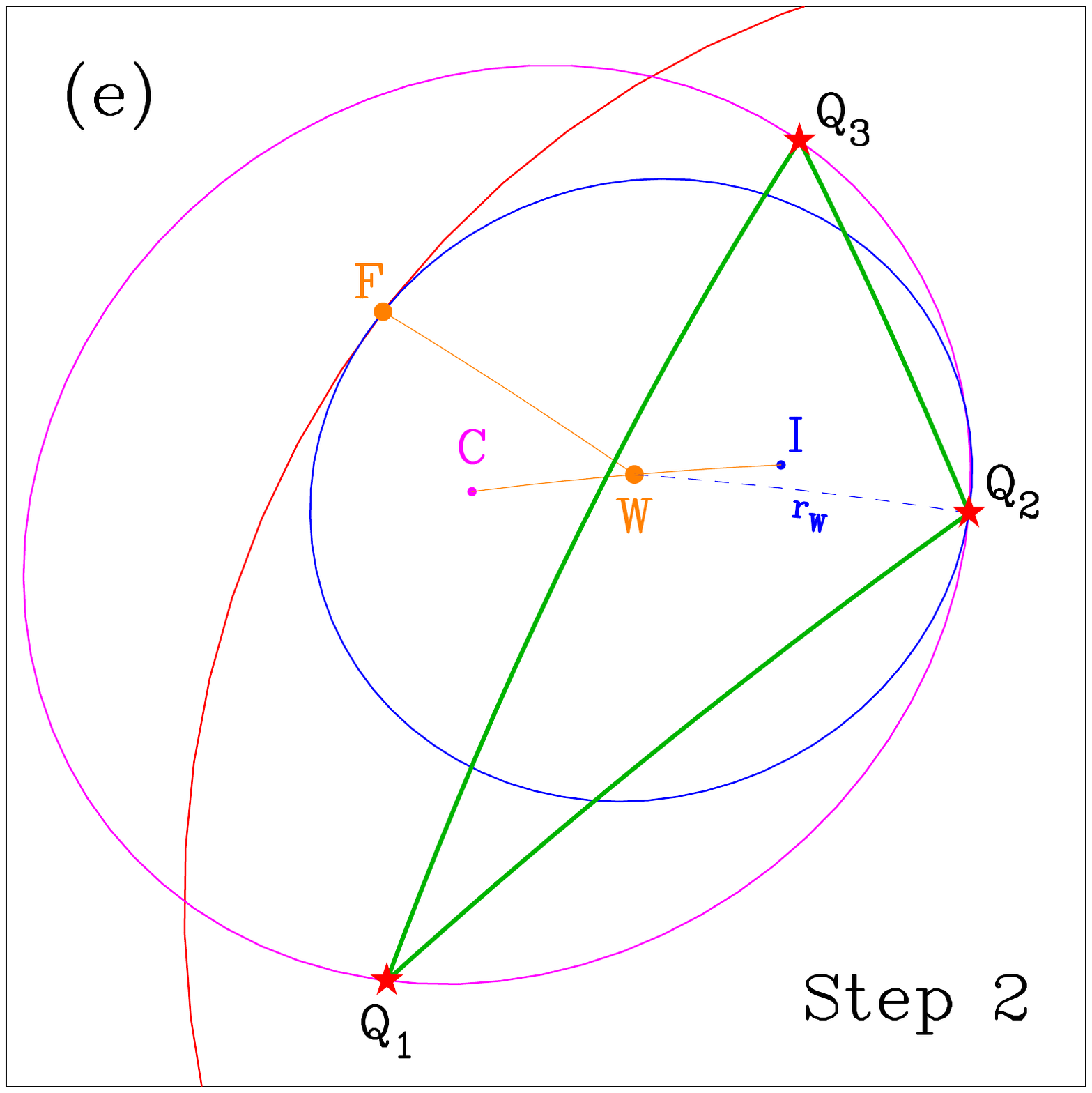}
\hfill
\includegraphics[viewport=201 79 641 518,width=0.66\columnwidth]{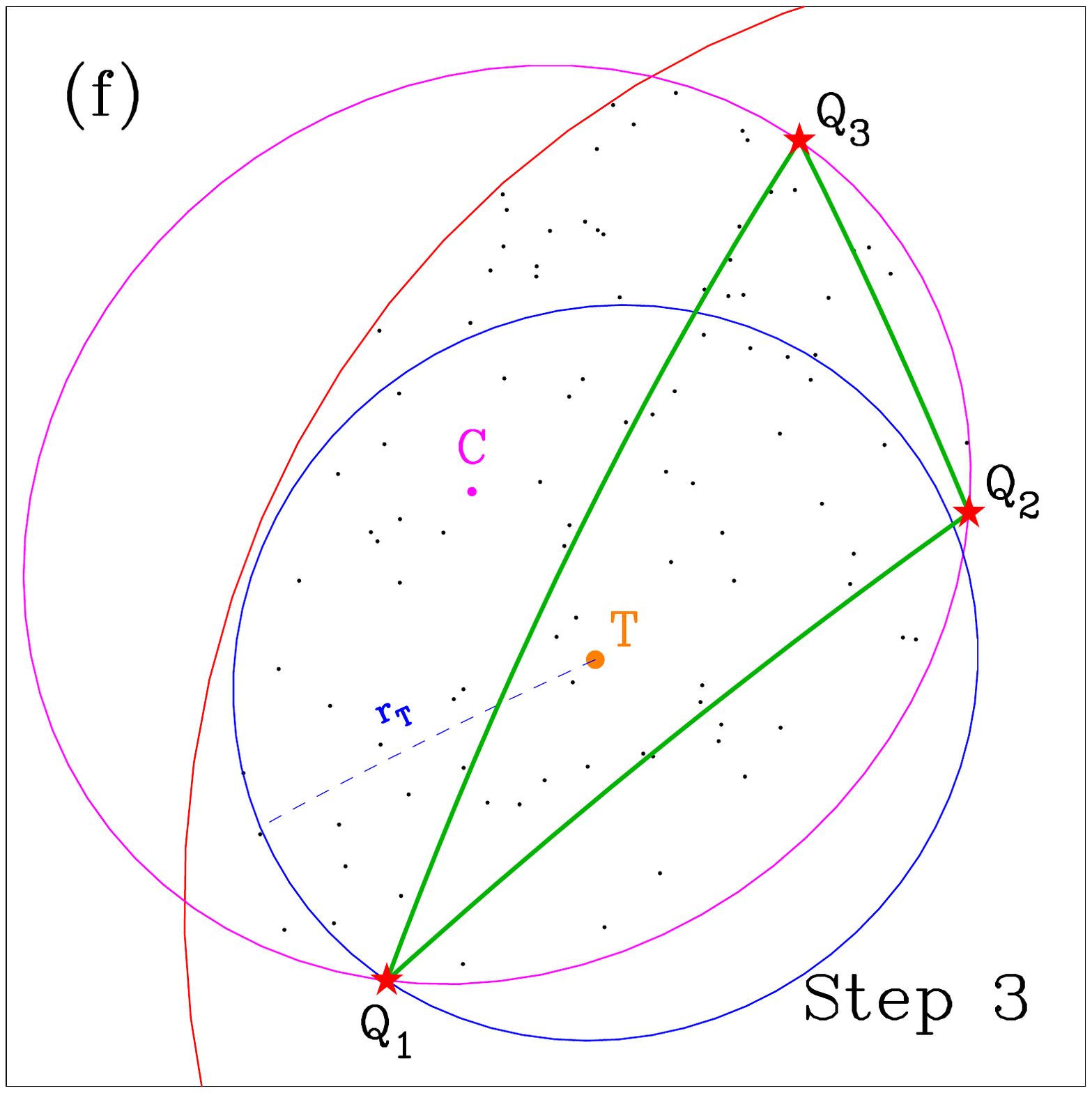}
\caption{Different strategies described in
Appendix~\ref{appendix:boundary-triangles} to obtain a valid blank field in the
neighborhood of a boundry triangle. These diagrams represent an enlargement of
the example represented in Fig.~\ref{figure:step0}, where, for clarity, some
graphical elements represented in that figure have been removed.\newline
{--- \bf Diagram (a)}: Computation of the incircle, the circle inscribed in the
spherical triangle. It is centered at $\bmath{I}$ and is displayed in blue. The
distance from $\bmath{I}$ to each of the sides of the boundary triangle is the
same, $r_I$.\newline
{--- \bf Diagram (b)}: Enlargement of the initial incircle until reaching the
closest star in the boundary triangle. Note that the three stars at the
vertices of the spherical triangle have been renamed as $\bmath{Q_\dd{1st}}$,
$\bmath{Q_\dd{2nd}}$ and~$\bmath{Q_\dd{3rd}}$ to indicate their order of
proximity to the incenter. Thus the resulting new radius,
$r_{I,\dd{enlarged}}$, is the angular distance between the incircle $\bmath{I}$
and $\bmath{Q_\dd{1st}}$.\newline
{--- \bf Diagram (c)}:
Displacement of the center of the blank field along the great circle
connecting the incenter $\bmath{I}$ with the second nearest triangle node
$\bmath{Q_\dd{2nd}}$. The new center $\bmath{V}$ is equidistant to the nodes
$\bmath{Q_\dd{1st}}$ and~$\bmath{Q_\dd{2nd}}$.\newline
{--- \bf Diagram (d)}:
Displacement of the center of the blank field along the great circle
connecting the incenter $\bmath{I}$ with the furthest triangle node
$\bmath{Q_\dd{3rd}}$. The new center $\bmath{V}$ is equidistant to the nodes
$\bmath{Q_\dd{1st}}$ and~$\bmath{Q_\dd{3rd}}$.\newline
{--- \bf Diagram (e)}:
Displacement of the center of the blank field along the great circle
connecting the incenter $\bmath{I}$ and the circumcenter $\bmath{C}$. The
distance $r_W$ from the new center $\bmath{W}$ to the nearest triangle star is
the same than the distance to the boundary of the sky subregion, indicated by
$\bmath{F}$.\newline
{--- \bf Diagram (f)}:
Random search of a valid blank field around the spherical triangle
defined by $\bmath{Q_1}$, $\bmath{Q_2}$ and~$\bmath{Q_3}$. The black dots
are random points in the intersection of the circumcircle and the sky
subregion. Each of these points is checked as the center of a potential blank 
field. The blue circle, with center $\bmath{T}$ and radius $r_T$, indicates the
largest of such circles that satisfy the conditions for a valid blank field.
}
\label{figure:steps}
\end{figure*}

\subsection*{\emph{Discovering problematic boundary circumcircles}}

The Delaunay triangulation provides the coordinates of the three nodes (stars)
of all the triangles resulting from the tessellation procedure (see
Fig.~\ref{figure:step0}). Let denote the unit vectors pointing from $\bmath{O}$
toward the three nodes of each triangle, given in counterclockwise order, as
$\bmath{Q_1}, \bmath{Q_2}, \bmath{Q_3}$, The unit vector $\bmath{C}$ marking
the circumcentre of any triangle can be easily computed as \citep[see
e.g.][]{renka97}
\begin{equation}
\bmath{C} = \frac{(\bmath{Q_2}-\bmath{Q_1}) \times (\bmath{Q_3}-\bmath{Q_1})}%
{||(\bmath{Q_2}-\bmath{Q_1}) \times (\bmath{Q_3}-\bmath{Q_1})||}.
\label{eq:circumcenter}
\end{equation}
Note that since the three nodes define a plane, the two vectors
\mbox{$(\bmath{Q_2}-\bmath{Q_1})$} and \mbox{$(\bmath{Q_3}-\bmath{Q_1})$} are
contained in that plane, and their cross product provides a new vector
perpendicular to the plane. Of the two possible directions of that vector, the
fact that the nodes are given in counterclockwise order guarantees that
$\bmath{C}$ points toward the spherical triangle defined by $\bmath{Q_1}$,
$\bmath{Q_2}$, and $\bmath{Q_3}$. In addition, the modulus calculated in the
denominator of Eq.~\ref{eq:circumcenter} assures that
$\bmath{C}$ is normalised. It is not difficult to show that the
circumcentre is equidistant from all the three nodes, which allows to compute
the circumradius of the circumcircle from the inner product of $\bmath{C}$ with
any of the node vectors
\begin{equation}
\bmath{r_C} = \arccos(\bmath{C} \cdot \bmath{Q_i}), \qquad 
\mbox{with $i=1$, 2 or 3.}
\label{eq:circumradius}
\end{equation}
On the other hand, the angular distance between the circumcenter and the center
of the sky subregion can be determined as
\[
d_C = \arccos(\bmath{C} \cdot \bmath{K_\dd{sky}}).
\]
The condition that must be met for any circumcircle to be inscribed within that
subregion is then
\begin{equation}
d_C+ r_C \leq \theta_\dd{max}.
\label{equation:condition-inside}
\end{equation}
If the above condition does not hold, the computed circumcircle expands beyond
the sky subregion and cannot be used as a valid blank field. Under this
circumstance other options need to be considered, as explained in the
next steps.

\subsection*{\emph{Step 0: compute the circle inscribed in the triangle}}

No matter how close to the boundary of the sky subregion is located any
triangle computed during the tessellation, the circle inscribed in those
triangles are always within the subregion. These \emph{incircles} are centered
at a particular point, the \emph{incenter}, which is equidistant from the three
sides of the triangles and that corresponds to the intersection of the three
bisectors through the vertices of those triangles (see
Fig.~\ref{figure:steps}a). The unit vector~$\bmath{I}$ pointing
from~$\bmath{O}$ toward the incenter of a given triangle can be obtained as
\begin{equation}
\bmath{I} = \frac{(\bmath{N_{1,2}}-\bmath{N_{3,1}}) \times 
                  (\bmath{N_{2,3}}-\bmath{N_{1,2}})}%
{||(\bmath{N_{1,2}}-\bmath{N_{3,1}}) \times 
(\bmath{N_{2,3}}-\bmath{N_{1,2}})||},
\end{equation}
where $\bmath{N_{i,j}}$ is the unitary vector perpendicular to the plane
defined by the origin and the nodes at $\bmath{Q_i}$ and~$\bmath{Q_j}$,
\begin{equation}
\bmath{N_{i,j}} = \frac{\bmath{Q_i} \times \bmath{Q_j}}%
{||\bmath{Q_i} \times \bmath{Q_j}||}.
\end{equation}
Defined in this way, and considering that the three triangle nodes are given in
counterclockwise order, the three normal vectors $\bmath{N_{1,2}}$,
$\bmath{N_{2,3}}$ and $\bmath{N_{3,1}}$ are always pointing toward the
interior of the triangles.
It is also easy to show that the inner product
\mbox{$\bmath{I} \cdot \bmath{N_{i,j}}$} provides the radius of the
incircle as
\begin{equation}
r_I = \arcsin(\bmath{I} \cdot \bmath{N_{i,j}}), 
\label{equation:radius-incenter}
\end{equation}
whose value is the same for $(i,j)=(1,2)$, $(2,3)$ or~$(3,1)$. Note 
the use of
the $\arcsin$ function instead of the $\arccos$ in the previous expression;
since $\bmath{N_{i,j}}$ is perpendicular to the plane defined by the origin
$\bmath{O}$ and the nodes at $\bmath{Q_i}$ and $\bmath{Q_j}$, the angular
distance $r_I$ between $\bmath{I}$ and that plane must be computed as $\pi/2$
minus the angle subtended by $\bmath{I}$ and $\bmath{N_{i,j}}$.

By construction it is obvious that incircles will always be regions devoided of
stars, since they are inscribed in the considered triangles, that in turn are
inscribed in the circumcircles, which, by the empty circumcircle interior
property (see Section~\ref{subsection:delaunay-triangulation}) do not contain
any star. For that reason they constitute a first solution to the problem with
the boundary triangles which circumcircles expand beyond the limit of the sky
subregion under tessellation.

\subsection*{\emph{Step 1: enlarge the radius toward the triangle nodes}}

Once the incircle of a given boundary triangle has been computed, it is very
likely that the incenter does not coincide with the circumcenter. In this case
it is possible to enlarge the incircle by increasing the radius $r_I$ to the
closest triangle node as measured from the incenter (see
Fig.~\ref{figure:steps}b). To facilitate the explanation of the
procedure, let rename the unit vectors pointing toward the three nodes of each
triangle as $\bmath{Q_\dd{1st}}$, $\bmath{Q_\dd{2nd}}$, and
$\bmath{Q_\dd{3rd}}$, where the subscript indicates the ordering of the nodes
as a function of their proximity to the incenter.

The enlargement of the angular distance connecting $\bmath{I}$ with
$\bmath{Q_\dd{1st}}$ provides a new radius than can be expressed as
\begin{equation}
r_{I,\dd{enlarged}} = 
\mbox{arccos}(\bmath{I} \cdot \bmath{Q_\dd{1st}}).
\label{equation:radius-incenter-enlarged}
\end{equation}
However, although the the new enlarged circle will only pass through one of the
nodes and leave outside the other two nodes, it is possible that other
neighbouring stars can enter into its interior. For that reason, instead of
Eq.~\ref{equation:radius-incenter-enlarged} it is necessary to use
\begin{equation}
r_{I,\dd{enlarged}} = 
\mbox{minimum}\left[ \mbox{arccos}(\bmath{I} \cdot \bmath{S_i})\right], 
\,\, \forall i=1,\ldots,N_{\dd{stars}},
\label{equation:radius-incenter-enlarged-final}
\end{equation}
which already includes the relation given in
Eq.~\ref{equation:radius-incenter-enlarged}, since the three nodes are part ot
the set of stars.

Before accepting $r_{I,\dd{enlarged}}$ as a valid radius for a blank field, it
is required to check that with this new radius the resulting circle (centered
at the triangle incenter) does not exceed the boundary of the sky subregion.
This is performed in a similar way as previously employed in step~0.  First the
angular distance between the incenter and the center of the sky subregion is
determined as
\[
d_I = \arccos(\bmath{I} \cdot \bmath{K_\dd{sky}}).
\]
The condition that must be met by the enlarged circle is then
\begin{equation}
d_I+ r_{I,\dd{enlarged}} \leq \theta_\dd{max}.
\end{equation}
If the above condition is not satisfied, a revised version of the radius can be
computed as
\begin{equation}
r_{I,\dd{enlarged},\dd{revised}} = \theta_\dd{max}-d_I.
\end{equation}

Note that, so far, the valid blank field is still centered around the triangle
incenter. However, by removing this constraint, it is possible to obtain
an even larger valid blank field by allowing its center to move from the
incenter toward the directions of the other two nodes following the great
circles connecting these points. 

Next we describe the procedure of shifting the centre of the
blank field along the line connecting the incenter with the node
corresponding to $\bmath{Q_\dd{2nd}}$. The resulting new center
$\bmath{V}$ (see Fig.~\ref{figure:steps}c) will be a point equidistant
from $\bmath{Q_\dd{1st}}$ and $\bmath{Q_\dd{2nd}}$. The situation with
the third node will be identical interchanging $\bmath{Q_\dd{2nd}}$ by
$\bmath{Q_\dd{3rd}}$ in the following description (see
Fig.~\ref{figure:steps}d).

The first step consists in computing the midpoint in the triangle side
connecting $\bmath{Q_\dd{1st}}$ and $\bmath{Q_\dd{2nd}}$. The unitary
vector pointing toward this point is given by
\begin{equation}
\bmath{M} = \frac{\bmath{Q_\dd{1st}}+\bmath{Q_\dd{2nd}}}%
{||\bmath{Q_\dd{1st}}+\bmath{Q_\dd{2nd}}||}.
\end{equation}
The sought new center $\bmath{V}$ can be obtained as the intersection of the
great circle defined by two vectors $\bmath{M}$ and $\bmath{C}$ with the great
circle defined by the two vectors $\bmath{I}$ and $\bmath{Q_\dd{2nd}}$. This
intersection can be easily computed as
\begin{equation}
\bmath{V} = \pm
\frac{(\bmath{C}\times\bmath{M}) \times (\bmath{I}\times\bmath{Q_\dd{2nd}})}%
{||(\bmath{C}\times\bmath{M}) \times (\bmath{I}\times\bmath{Q_\dd{2nd}})||},
\end{equation}
where the $\pm$ sign indicates two initial solutions from which the closest
vector to $\bmath{I}$ must be selected. The angular distance from the new
center $\bmath{V}$ to either $\bmath{Q_\dd{1st}}$ and $\bmath{Q_\dd{2nd}}$ is 
now
\begin{equation}
r_V = \arccos(\bmath{V} \cdot \bmath{Q_\dd{node}}), \quad
\mbox{with node=1st or 2nd}.
\end{equation}
Next, it is necessary to check that no new stars (different from the
ones at the triangle nodes) have entered within the new circle. This can be
computed as
\begin{equation}
r_{V,\dd{revised}} = 
\mbox{maximum} \left\{ 
r_V, \mbox{minimum}[\arccos(\bmath{V} \cdot \bmath{S_i})] 
\right\},
\end{equation}
where $i=1,\ldots,N_\dd{stars}$.

Note that it must also be checked that with this new radius
$r_{V,\dd{revised}}$ the circle centered at $\bmath{V}$ is still within the sky
subregion. In a similar same as we proceeded in Step~2, the distance from
$\bmath{V}$ to $\bmath{K_\dd{sky}}$ is determined as
\begin{equation}
d_V = \arccos(\bmath{V} \cdot \bmath{K_\dd{sky}}),
\end{equation}
and then the condition that must be satified can be written as
\begin{equation}
d_V + r_{V,\dd{revised}} \leq \theta_\dd{max}.
\end{equation}
If the previous condition does not hold, the radius of the blank field can be
redefined as
\begin{equation}
r_{V,\dd{revised2}} = \theta_\dd{max}-d_V.
\end{equation}

The resulting blank field centered at $\bmath{V}$ will be a better (larger)
blank field as far as $r_{V,\dd{revised}}$ (or $r_{V,\dd{revised2}}$)
is larger than $r_{I,\dd{enlarged}}$.

\subsection*{\emph{Step 2: shift the center between the incenter
and the circumcenter}}

Independently of the success of the previous step, another possibility
worth being explored is shifting the center of the blank field along the
great circle connecting the incenter with the circumcenter, with the constraint
that the new circle does not cross the sky subregion boundary (see
Fig.~\ref{figure:steps}e).

For this computation the bisection method can easily be employed. Since the
solution for the new center will be a point in the great circle connecting
$\bmath{I}$ with $\bmath{C}$, it is possible to use two auxiliary vectors
$\bmath{L_1}$ and~$\bmath{L_2}$ which are initialized as
\mbox{$\bmath{L_1}=\bmath{I}$} and \mbox{$\bmath{L_2}=\bmath{C}$}. Then the
process enters into an iterative procedure in which the following steps
are performed:
\begin{enumerate}

  \item The midpoint on the great circle connecting $\bmath{L_1}$
  and~$\bmath{L_2}$ is determined as
  \[
     \bmath{L_\dd{mid}} =
     \frac{\bmath{L_1}+\bmath{L_2}}{||\bmath{L_1}+\bmath{L_2}||}.
  \]

  \item The minimum angular distance from the previous point to any of the 
  triangle nodes is evaluated using
  \[
      d_\dd{L--Q} = \mbox{minimum}  
        [ \arccos( \bmath{L_\dd{mid}} \cdot \bmath{Q_i} ) ], \quad
      \mbox{with $i=1$, 2 or 3.}
  \]

  \item The angular distance from $\bmath{L_\dd{mid}}$ to the border of the sky
  subregion is also determined as
  \[
      d_\dd{L--Sky} = 
        \theta_\dd{max} - \arccos(\bmath{L_\dd{mid}} \cdot \bmath{K_\dd{sky}}).
  \]

  \item If the two previous angular distances are equal within a given
  tolerance (for our purposes $|d_\dd{L--Q}-d_\dd{L--Sky}| < 0.01$~arcsec is
  enough), the iterative procedure halts and the final result is defined as
  \mbox{$\bmath{W}=\bmath{L_\dd{mid}}$}. Otherwise one of the auxiliary vectors
  must be redefined according to
  \[
      \mbox{if}\,\, (d_\dd{L--Q}-d_\dd{L--Sky}) \,\,\mbox{is}
      \left\{
      \begin{array}{l}
        > 0, \quad \mbox{then} \,\,\, \bmath{L_2} = \bmath{L_\dd{mid}} \\
        < 0, \quad \mbox{then} \,\,\, \bmath{L_1} = \bmath{L_\dd{mid}} \\
      \end{array}
      \right.
  \]
  and the process is iterated by repeating steps (i)--(iv).

\end{enumerate}

At the end of the iterative process a new center $\bmath{W}$ has been computed
(see Fig.~\ref{figure:steps}e), which is equidistant to the nearest triangle
node and to the boundary of the sky subregion, i.e.,
\mbox{$r_W=d_\dd{L--Sky}=d_\dd{L-Q}$}.

Finally it must be checked that no new stars have entered into the new circle.
Similarly to what was done in previous steps, the new radius can be refined by
using
\begin{equation}
r_{W,\dd{revised}} = 
\mbox{maximum}\{r_W, \mbox{minimum}[\arccos(\bmath{W} \cdot \bmath{S_i})] \}
\end{equation}
where $i=1,\ldots,N_\dd{stars}$.

It is important to highlight that new radius $r_{W,\dd{revised}}$ will not
necessarily be larger than the radii previously derived in Steps~2 and~3 (in
fact, the example illustrating this appendix shows that the blank field in
Fig.~\ref{figure:steps}d centered at $\bmath{V}$ with radius $r_V$ is larger
than the blank field in Fig.~\ref{figure:steps}e centered at $\bmath{W}$ with
radius $r_W$).

\subsection*{\emph{Step 3: random search in the intersection between the
circumcircle and the sky subregion}}

A final, but also effective, brute force method consists in using a random
search of a valid blank field center within the intersection region between the
circumcircle and the sky subregion (see Fig.~\ref{figure:steps}f).  For
this purpose, random points are generated on that region, imposing that any
small area has to contain, on average, the same number of points \citep[see
e.g.][]{weisstein11}. For each random point $\bmath{T}$ the valid blank field
radius is evaluated as
\begin{equation}
r_T = \mbox{minimum}[\arccos(\bmath{T} \cdot \bmath{S_i})], 
\quad \forall i=1,\ldots,N_\dd{stars}.
\end{equation}
It is also necessary to check that the resulting blank field remains inscribed
in the sky subregion. For that purpose the distance from the center to the
border of the subregion is determined using
\begin{equation}
d_T=\arccos(\bmath{T} \cdot \bmath{K_\dd{sky}}),
\end{equation}
and the corresponding condition to be verified is
\begin{equation}
d_T + r_T \leq \theta_\dd{max}.
\end{equation}
If this is not the case, the value of the radius can be refined using
\begin{equation}
r_{T,\dd{refined}} = \theta_\dd{max}-d_T.
\end{equation}

Obviously, the solution obtained with this method must be compared with the
solutions derived in the previous steps in order to decide which one provides
the largest blank field in the neighborhood of the considered boundary
triangle.

\subsection*{\emph{Successfulness of the different steps}}

The three steps just described produce different possible valid blank fields
associated to each boundary triangle. The adopted valid blank field in each
case will be the largest among those alternatives. 

\begin{figure}
\centering
\includegraphics[viewport=40 106 743 562,width=\columnwidth]{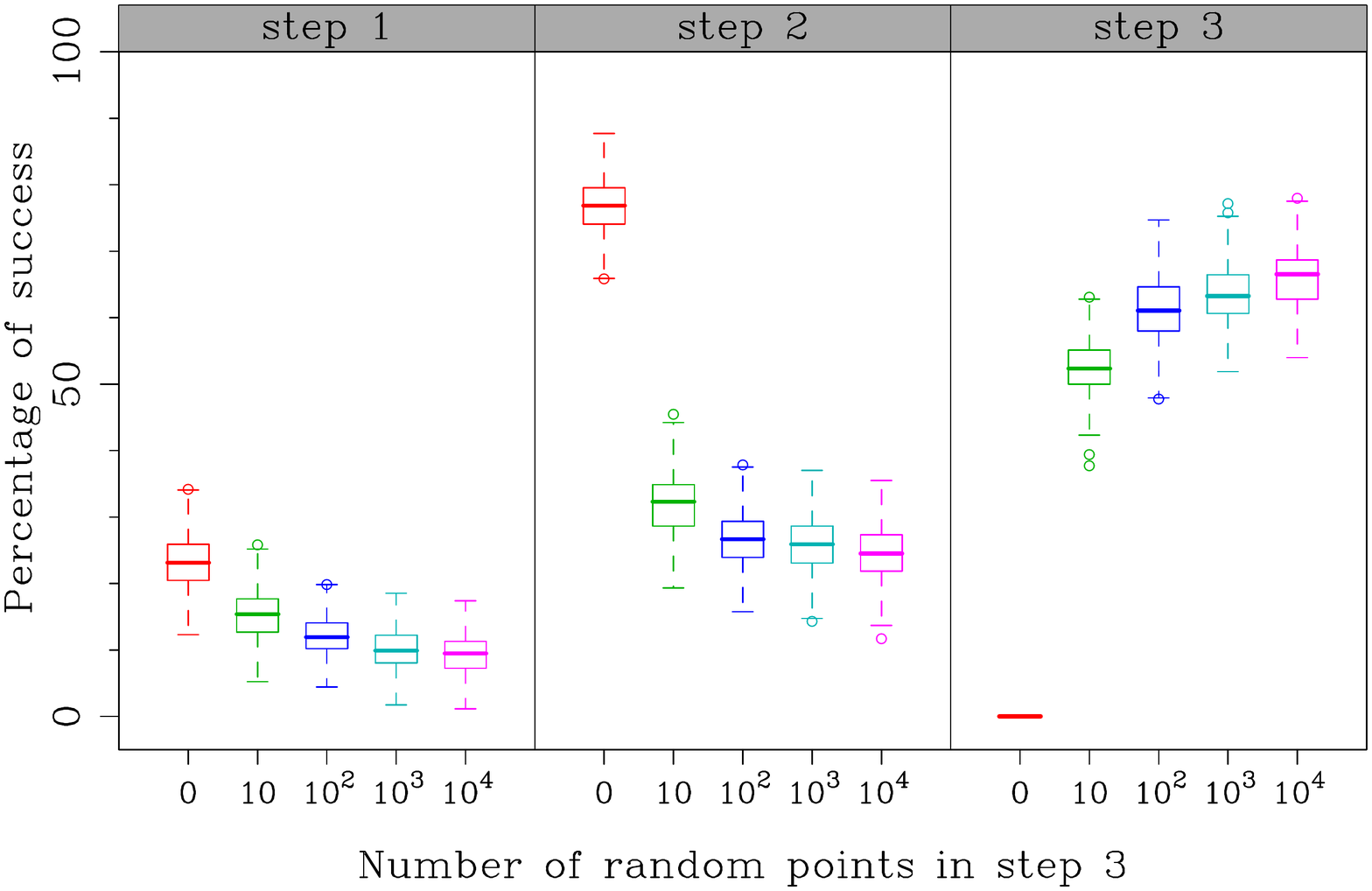}

\vskip 5mm

\caption{Box-and-whisker plot showing the percentage of success of the
different steps described in Appendix~\ref{appendix:boundary-triangles}. The
bottom and top of each box show the 25th and 75th percentiles, whereas the
thick horizontal line marks the median. The whiskers (vertical dashed lines)
show 1.5~times the interquartile range. Since each colour represents a fixed
number of random points employed in the search for valid a blank field region
in step~3, the sum of the percentages in the three steps for a fixed colour
is~100.}
\label{stepspercent}
\end{figure}

It is interesting to investigate the successfullness of the three steps
providing the largest blank field. As a first-order estimation of these
numbers, we have measured the percentage of success of each step when
tessellating random stellar fields. Since the success ratio for step~3 depends
on the number of random points drawn within the intersection region between the
circumcircle and the sky subregion, we have computed those percentages using
different numbers of random points. The results are displayed in
Fig.~\ref{stepspercent}. Each colour in this figure indicates a fixed number of
random points employed in step~3. When no random points are used (represented
in red colour), the median percentage of success of steps~1 and~2 are 23\% and
77\%, respectively. Not surprisingly these values decrease as the number of
random points in step~3 increases, reaching a stable situation as far as the
number of random points exceeds a few hundred. In particular, for $10^4$ random
points the median percentages of success (represented in magenta colour) are
10\%, 24\% and 66\% for steps 1, 2 and~3, respectively. 

The previous results have shown that although step~3 will be, in general, the
resposible for providing the valid blank field, the contribution of the other
two steps is not negligible.

\label{lastpage}

\end{document}